\documentclass[final,twocolumn,5p,times,10pt,authoryear]{elsarticle}
\usepackage{lineno}
\modulolinenumbers[5]

\journal{arXiv}

\usepackage[american]{babel}
\usepackage[utf8]{inputenc}
\usepackage[T1]{fontenc}
\usepackage{graphicx}
\usepackage{amsmath}
\usepackage{amssymb}
\usepackage{amsthm}
\usepackage{amsfonts}
\usepackage{bm}
\usepackage{hhline}
\usepackage{csquotes}
\usepackage{color}
\usepackage{xcolor}
\usepackage{import}
\usepackage{algorithm}
\usepackage{algpseudocode}
\usepackage{dsfont}
\usepackage{multirow}
\usepackage{soul}
\setstcolor{magenta}

\makeatletter
\usepackage[colorlinks=true,linkcolor=red]{hyperref}
\makeatother

\newtheorem{theorem}{\bf{Theorem}}
\newtheorem{lemma}{\bf{Lemma}}
\newtheorem{proposition}{\bf{Proposition}}
\newtheorem{definition}{\bf{Definition}}
\newtheorem{remark}{\bf{Remark}}
\newtheorem{corollary}{\bf Corollary}
\newtheorem{assumption}{\bf Assumption}

\newcommand{\mbf}[1]{\ensuremath{{\mathbf{#1}}}}

\newcommand{\ten}[1]{\ensuremath{{\cdot}10^{#1}}}

\newcommand{\half}{\ensuremath{\frac{1}{2}}}

\newcommand{\eye}[1]{\ensuremath{\mbf{I}_{#1}}}

\newcommand{\zeros}[2]{\ensuremath{\bm{0}_{#1\times#2}}}
\newcommand{\ones}[2]{\ensuremath{\bm{1}_{#1\times#2}}}

\newcommand{\real}[1]{\ensuremath{\text{Re}(#1)}}

\newcommand{\realset}{\ensuremath{\mathbb{R}}}
\newcommand{\realsetmat}[2]{\ensuremath{\mathbb{R}^{#1\times#2}}}
\newcommand{\intvalset}{\ensuremath{\mathbb{I}\mathbb{R}}}

\newcommand{\lbound}{\ensuremath{\text{L}}}
\newcommand{\ubound}{\ensuremath{\text{U}}}
\newcommand{\midpoint}{\ensuremath{\text{M}}}

\newcommand{\radius}{\ensuremath{\text{R}}}

\newcommand{\lb}[1]{{#1}^\lbound}
\newcommand{\ub}[1]{{#1}^\ubound}
\newcommand{\midp}[1]{{#1}^\midpoint}
\newcommand{\rad}[1]{{#1}^\radius}

\newcommand{\poly}{\ensuremath{_\text{P}}}
\newcommand{\zon}{\ensuremath{_\text{Z}}}
\newcommand{\czon}{\ensuremath{_\text{CZ}}}

\newcommand{\intval}[1]{\ensuremath{[\lb{#1}, \ub{#1}]}}

\newcommand{\ninf}[1]{\ensuremath{\|{#1}\|_\infty}}

\newcommand{\noarg}{\ensuremath{\_\,}}

\newcommand{\intvol}[1]{\ensuremath{\text{Vol}_\square(#1)}}
\newcommand{\parvol}[1]{\ensuremath{\text{Vol}_\diamond(#1)}}
\newcommand{\intavr}{\ensuremath{\text{GAVR}_\square}}
\newcommand{\paravr}{\ensuremath{\text{GAVR}_\diamond}}

\newcommand{\betacv}{\ensuremath{\beta^\text{CV}_{j}}}
\newcommand{\betacc}{\ensuremath{\beta^\text{CC}_{j}}}

\newcommand{\code}[1]{\texttt{\detokenize{#1}}}

\newcommand{\bibfolder}{Bibliography}

\biboptions{authoryear}

\allowdisplaybreaks 

\newif\ifarxiv

\begin{document}

\begin{frontmatter}

\title{Set-based state estimation of nonlinear discrete-time systems using constrained zonotopes and polyhedral relaxations\tnoteref{mytitlenote}}

\tnotetext[mytitlenote]{This work was partially supported by the Brazilian agencies CNPq, under grants 465755/2014-3 (INCT project), and 315695/2020-0; FAPESP, under the grants 2014/50851-0, 2022/05052-8 and 2023/06896-8; CAPES, under the grants 001 and 88887.136349/2017-00; and FAPEMIG, under grant APQ-03090-17.}

\author[usp]{Brenner S. Rego\corref{mycorrespondingauthor}}
\cortext[mycorrespondingauthor]{Corresponding author, \textit{brennersr7@usp.br} }

\author[ufmg]{Guilherme V. Raffo}

\author[usp]{Marco H. Terra}

\author[georgiatech]{Joseph K. Scott}

\address[usp]{Department of Electrical and Computer Engineering, University of São Paulo, São Carlos, SP 13566-590, Brazil}
\address[ufmg]{Department of Electronics Engineering, Federal University of Minas Gerais, Belo Horizonte, MG 31270-901, Brazil}
\address[georgiatech]{Department of Chemical and Biomolecular Engineering, Georgia Institute of Technology, 311 Ferst Dr., Atlanta, 30318, GA, USA}

\begin{abstract}
This paper presents a new algorithm for set-based state estimation of nonlinear discrete-time systems. A key step in such algorithms is to propagate a set (often a zonotope or constrained zonotope) through a nonlinear function. Existing methods accomplish this through conservative linearization procedures that are known to lead to severe overestimation in many cases. Here, we propose an alternative that avoids linearization using the so-called factorable representation of nonlinear functions. We use a recursive polyhedral relaxation technique based on this representation that is well-established in the global optimization literature but has not previously been used for set-based estimation. This technique is combined with constrained zonotope (CZ) technology to avoid the limitations of recursive computations with polyhedra in halfspace representation. The resulting state estimation method is fully automated, has attractive computational complexity (with one caveat discussed herein), and can provide significantly tighter enclosures than those resulting from linearization procedures in many cases. Numerical examples highlight the advantages of this approach relative to existing CZ methods based on the Mean Value Theorem and Difference of Convex functions (DC) programming.
\end{abstract}



\end{frontmatter}


\section{Introduction}

This paper develops a new method for set-based state estimation of nonlinear discrete-time systems with bounded uncertainties. Set-based computations are an important topic in the recent literature, with several applications to dynamical systems, such as reachability analysis \citep{Yang2020,Althoff2021Reach}, state estimation \citep{Rego2021}, parameter identification \citep{Wang2017ZonotopeParameter}, joint state-parameter estimation \citep{Rego2022Joint}, fault diagnosis \citep{Raimondo2016,Mu2024FDD}, and tube-based model predictive control \citep{Zhang2022KoopmanMPC,AND2024}. In particular, set-based state estimation generates guaranteed enclosures of the system trajectories consistent with the measured outputs, which is especially important in mobile robotics \citep{Colle2019SetMobileRobot}, aerospace systems \citep{Wang2025SatelliteZonotope}, power systems \citep{Li2025PowerSystems}, and more.

Set-based state estimation algorithms require the ability to propagate sets through the functions defining the system's dynamics and measurements (i.e., to compute enclosures of the possible values of these functions over given sets of inputs). The literature has well established solutions for set-based state estimation of linear systems \citep{Althoff2021Comparison}. However, when these functions are nonlinear, computing tight enclosures while maintaining reasonable computational complexity remains a significant challenge. Additionally, such computations often lead to an increase in the complexity of the resulting enclosure at each time step (i.e., the size of the data needed to represent the set computationally). This increase can be linear, quadratic, or even exponential, and can become prohibitive even over relatively short horizons. Thus, complexity reduction algorithms, which compute an enclosure of a given set with lower complexity, have become another critical component of set-based estimation methods.

To address these challenges, state-of-the-art methods often represent enclosures as zonotopes \citep{Alamo2005a,Pan2023Koopman} or one of their generalizations, including constrained zonotopes (CZs) \citep{Rego2021}, constrained polynomial zonotopes \citep{Kochdumper2023CPZ}, and hybrid zonotopes \citep{Siefert2025HZFuncDecomp}. This work focuses on zonotopes and CZs due to their computational advantages for some important set operations and the availability of effective complexity reduction algorithms with polynomial-time complexity \citep{Scott2016,Scott2018,Raghuraman2022}. 
In most existing estimation methods based on zonotopes and CZs, the crucial set propagation steps are done using linear approximations of the dynamics and measurement equations. The resulting enclosures are then made rigorous by adding bounds on the linearization errors obtained using the Mean Value Theorem and Taylor's Theorem \citep{Alamo2005a,Combastel2005,Rego2021}, which lead to linear and quadratic growth of the set complexity, respectively. For small uncertainties, these linear approximations may result in good enclosures. However, because the linearization error must be bounded over the entire set of inputs to the function, severe conservatism is often observed when the input set is large. In the propagation from time $k-1$ to $k$, this input set is the joint set of possible uncertainties and states at time $k-1$, which can be large even if it is an accurate enclosure at $k-1$. Thus, the only generally effective way to manage this limitation is through state-space partitioning, which adds prohibitive costs.

As an alternative to the Mean Value Theorem and Taylor's Theorem, Difference of Convex functions (DC) programming principles have been used to obtain improved bounds on the linearization error \citep{Alamo2008,DePaula2024DC}. However, despite the linear growth in set complexity achieved by these algorithms, their computational costs scale exponentially with the number of states, and the resulting enclosures may not be accurate for some functional forms \citep{Tottoli2023}. Additional zonotope methods based on the Koopman operator have also been proposed \citep{Pan2023Koopman}. These methods also achieve linear growth in set complexity and do not require any linearization error bounds. Instead, they rely on nonlinear variable transformations that map the system into a lifted space where the dynamics and measurement equations are linear. However, this requires propagating the initial sets through these nonlinear transformations, which is often difficult. Existing methods use non-rigorous affine approximations obtained from a least-squares solution for samples that are generated from the initial sets, which may not result in guaranteed enclosures. Besides, except for convenient functional forms, this approach may require transformation into an infinite-dimensional space to obtain good approximations.

In this work, we develop a novel method for state estimation of nonlinear discrete-time systems using constrained zonotopes and polyhedral relaxations. This method, referred to as CZPR, allows for the propagation of CZs through nonlinear dynamics and measurement equations without the need for linearization and the computation of linearization errors. This is accomplished by adapting a method commonly used in the global optimization literature for computing linear programming relaxations of nonlinear optimization problems \citep{Tawarmalani2002Convexification,Tawarmalani2005Polyhedral}. In this method, nonlinear functions are decomposed into \emph{factorable representations} consisting of sequences of elementary operations. Enclosing the graph of each of these operations by a convex polyhedron allows for the generation of a polyhedral relaxation in a lifted space. Here, we adapt this technique to build a polyhedral enclosure of the image of a CZ input set under a composite function involving the system dynamics and measurement equations. Thereafter, the measured outputs are incorporated into the polyhedral enclosure through equality constraints. Properties of CZs are then employed to efficiently project the enclosure into the space of states at the next time step. This results in a recursive state estimation method with linear growth in the complexity of the CZs at each time step. This growth is addressed using polynomial-time complexity reduction algorithms for CZs from \cite{Scott2016}. 

The method proposed in this paper carries some similarities with \cite{SiefertHZEstimation2023} and \cite{Siefert2025HZFuncDecomp}, while having many differences. Factorable representations are a special case of the functional decompositions in \cite{Siefert2025HZFuncDecomp}, while the enclosures constructed from the decompositions are obtained using very different methods. The key difference relies in the way our method encloses the basic operations in the functional decomposition, which require only local bounds instead of global domains, and directly leads to a different choice of sets (CZs versus hybrid zonotopes) along with major implications for the rest of the algorithm. Computational results show that our approach can provide significantly better performance than CZ methods based on the Mean Value Theorem \citep{Rego2021} and DC programming \citep{DePaula2024DC} for some challenging examples.

Our previous conference paper \citep{Rego2024CZPRCDC} presented a preliminary version of this approach for reachability analysis. This paper extends those results in the following ways:
\begin{itemize}
    \item The CZ propagation method is extended to incorporate measured outputs, which enables state estimation as opposed to only reachability analysis;
    \item Polyhedral relaxations of trigonometric functions are developed, which extend the method to handle essentially all nonlinear functions;
    \item A new method is developed for representing the CZ output of our propagation method with fewer generators and constraints, leading to lower computational cost with no added conservatism;
    \item A more sophisticated and clear formulation of the concepts and methods is presented;
    \item More detailed computational comparisons are presented on an extended set of test cases.
\end{itemize}

The remainder of the manuscript is organized as follows. Section \ref{sec:preliminaries} presents the state estimation problem and the necessary mathematical background. Section \ref{sec:liftedpolyhedron} describes a method to obtain a lifted halfspace enclosure of the image of an interval under a nonlinear function. Section \ref{sec:stateestimation} combines the lifted enclosure with CZ operations to develop the complete state estimation method. Section \ref{sec:results} shows numerical examples, and Section \ref{sec:conclusions} concludes the paper.

\section{Preliminaries and problem formulation} \label{sec:preliminaries}

This section introduces the concepts of interval arithmetic, factorable functions, convex polyhedra, and constrained zonotopes. The state estimation problem is also formalized.

\subsection{Interval arithmetic}
\label{sec:Interval arithmetic}
Let $\intvalset^n$ denote the set of all non-empty compact intervals in $\realset^n$. For \emph{endpoints} $\lb{\mbf{x}}, \ub{\mbf{x}} \in \realset^n$ with $\lb{\mbf{x}} \leq \ub{\mbf{x}}$, an \emph{interval} $X \in \intvalset^n$ is defined as $X \triangleq \{\mbf{x} \in \realset^n: \lb{\mbf{x}} \leq \mbf{x} \leq \ub{\mbf{x}}\} \triangleq \intval{\mbf{x}}$. Define $\text{mid}(X) \triangleq \half (\ub{\mbf{x}} + \lb{\mbf{x}}) \triangleq \midp{\mbf{x}}$, $\text{rad}(X) \triangleq \half (\ub{\mbf{x}} - \lb{\mbf{x}}) \triangleq \rad{\mbf{x}}$, and $B_\infty^n \triangleq [-\ones{n}{1}, \ones{n}{1}]$. Moreover, for any bounded $Z \subset \realset^n$, let $\square Z \in \intvalset^n$ denote the interval hull of $Z$.

Let $X \triangleq \intval{x} \in \intvalset$ and $W \triangleq \intval{w} \in \intvalset$. For any of the arithmetic operations $\odot\in\{+,-,\times,/\}$, define $X \odot W \triangleq \{ x \odot y : x \in X, \ w \in W \}$ (division is undefined if $0 \in W$). Similarly, for any $h : \realset \to \realset$, define $h(X)\triangleq \{ h(x) : x \in X\}$. For any $\odot\in\{+,-,\times,/\}$ and a large library of standard univariate functions $h$ (including powers, exponentials, logarithms, trigonometric functions, etc.), the images $X \odot W$ and $h(X)$ are intervals and simple formulas for computing them in terms of the input endpoints are given in \cite{Moore2009}.

\subsection{Factorable functions}
\label{sec:Factorable functions}
In this paper, we consider nonlinear functions that are \emph{factorable} as defined below. This definition requires a library $\mathcal{L}$ of \emph{univariate functions}, which typically contains functions in a standard math library in any programming language, such as $x^a$, $e^x$, $\ln(x)$, $\sin{x}$. The concept of factorable functions has been explored since \cite{McCormick1976}.  While being a special case of functional decompositions defined in \cite{Siefert2025HZFuncDecomp}, this well-established methodology underlies interval arithmetic \citep{Moore1966}, many methods for nonconvex global optimization \citep{Tawarmalani2002Convexification}, and several methods for continuous-time reachability analysis \citep{Scott2013ImprovedRelaxations,Sahlodin2011ConvexConcave}.

\begin{definition} \label{def:factorable} \rm A function $\bm{\varphi}: D\subseteq\realset^{n_s} \to \realset^{n_\varphi}$ is said to be \emph{factorable} if it can be expressed in terms of a finite sequence of \emph{factors} $\zeta_j:D\rightarrow\mathbb{R}$ with $j\in\{1,\ldots,n_z\}$ such that:
	\begin{enumerate}
		\item For each $j\leq n_s$, $\zeta_j(\mathbf{s}) = s_j$, $\forall\mbf{s} \in D$;
		\item For each $j>n_s$, either
		\begin{enumerate}
			\item $\zeta_j(\mbf{s}) \triangleq \zeta_a(\mbf{s})\odot \zeta_b(\mbf{s})$, $\forall\mbf{s} \in D$, where $a,b < j$ and $\odot\in\{+,-,\times,/\}$, or
			\item $\zeta_j(\mbf{s}) \triangleq \beta_j(\zeta_a(\mbf{s}))$, $\forall\mbf{s} \in D$, where $a < j$ and $\beta_j$ is a univariate function in $\mathcal{L}$;
		\end{enumerate}
		\item $\bm{\varphi}(\mbf{s}) = \mbf{E}_\varphi\boldsymbol{\zeta}(\mbf{s})$, where $\boldsymbol{\zeta}=(\zeta_1,\ldots,\zeta_{n_z})$ and $\mbf{E}_\varphi \in \realsetmat{n_\varphi}{n_z}$ is a matrix of zeros except for a single $1$ in each of its rows (i.e., each output in $\bm{\varphi}$ is an element of $\boldsymbol{\zeta}$).
	\end{enumerate}
\end{definition}

For example, the function
\begin{equation*}
    \varphi(\mbf{s}) \triangleq \frac{s_1}{s_2 s_3 \sqrt[3]{\frac{(s_2+s_3)}{3}}}
\end{equation*}
can be factorized as $\zeta_1(\mbf{s}) \triangleq s_1$, $\zeta_2(\mbf{s}) \triangleq s_2$, $\zeta_3(\mbf{s}) \triangleq s_3$,
\begin{align*}
	\zeta_4(\mbf{s}) & \triangleq \zeta_2(\mbf{s}) \zeta_3(\mbf{s}), & & \zeta_7(\mbf{s}) \triangleq \sqrt[3]{\zeta_6(\mbf{s})}, \\
	\zeta_5(\mbf{s}) & \triangleq \zeta_2(\mbf{s}) + \zeta_3(\mbf{s}), & & \zeta_8(\mbf{s}) \triangleq \zeta_4(\mbf{s}) \zeta_7(\mbf{s}), \\
	\zeta_6(\mbf{s}) & \triangleq \frac{\zeta_5(\mbf{s})}{3}, & & \zeta_9(\mbf{s}) \triangleq \frac{\zeta_1(\mbf{s})}{\zeta_8(\mbf{s})}.
\end{align*}
With these definitions, $\varphi(\mbf{s}) = [\zeros{1}{8} \,\; 1]\boldsymbol{\zeta}(\mbf{s})$, and hence $\varphi$ is factorable. It is noteworthy that the factorization of $\bm{\varphi}(\mbf{s})$ is not unique. Additionally, assuming factorability is not very restrictive since any function that can be written explicitly in computer code using a standard math library is factorable.

In Definition \ref{def:factorable}, each factor is defined by applying a simple mathematical operation to one or two previous factors. The following definition provides a convenient notation for referring to these individual operations apart from the rest of the function.

\begin{definition} \label{def:factorable step functions} \rm Let $\bm{\varphi}: D\subseteq\realset^{n_s} \to \realset^{n_\varphi}$ be a factorable function with factors $\boldsymbol{\zeta}=(\zeta_1,\ldots,\zeta_{n_z})$. For each $j\in\{1,\ldots,n_z\}$, let $\alpha_j:\realset^{n_z}\rightarrow\mathbb{R}$ denote the corresponding elementary operation. Specifically:
	\begin{enumerate}
		\item For each $j\leq n_s$, $\alpha_j(\mathbf{z}) \triangleq z_j$, $\forall\mbf{z} \in \realset^{n_z}$;
		\item For each $j>n_s$,
		\begin{enumerate}
			\item if $\zeta_j(\mbf{s}) = \zeta_a(\mbf{s})\odot \zeta_b(\mbf{s})$, then $\alpha_j(\mbf{z}) \triangleq z_a\odot z_b$, $\forall\mbf{z} \in \realset^{n_z}$,
			\item if $\zeta_j(\mbf{s}) = \beta_j(\zeta_a(\mbf{s}))$, then $\alpha_j(\mbf{z}) \triangleq \beta_j(z_a)$, $\forall\mbf{z} \in \realset^{n_z}$.
		\end{enumerate}
	\end{enumerate}
\end{definition}

Note that each $\alpha_j$ depends on at most two components of $\mbf{z}$ (with indices less than $j$) and is only written as a function of the full vector for convenience. With this notation, we always have $\zeta_j(\mbf{s})=\alpha_j(\boldsymbol{\zeta}(\mbf{s}))$, $\forall \mbf{s}\in D$. 

\subsection{Convex polyhedra and constrained zonotopes}

\emph{Convex polyhedra} are commonly represented as the intersection of halfspaces. To facilitate some important manipulations, we use a slightly modified halfspace representation (H-rep) that stores equality constraints separately from inequalities.

\begin{definition} \rm \label{def:hrep}
	A set $P \subset \realset^n$ is a \emph{convex polyhedron} in halfspace representation if there exists $(\mbf{H}_p, \mbf{k}_p, \mbf{A}_p, \mbf{b}_p) \in \realsetmat{n_h}{n} \times \realset^{n_h} \times \realsetmat{n_{c_p}}{n} \times \realset^{n_{c_p}}$ such that
	\begin{equation} \label{eq:defhrep}
		P = \{ \mbf{x} \in \realset^n : \mbf{H}_p \mbf{x} \leq \mbf{k}_p, \mbf{A}_p \mbf{x} = \mbf{b}_p \}.
	\end{equation}	
\end{definition}

\emph{Constrained zonotopes} are an extension of zonotopes that include linear equality constraints. This allows CZs to describe arbitrary bounded convex polyhedra while retaining many of the computational advantages of zonotopes.

\begin{definition} \rm \citep{Scott2016} \label{def:pre_czonotopes}
	A set $Z \subset \realset^n$ is a \emph{constrained zonotope} if there exists $(\mbf{G}_z,\mbf{c}_z,\mbf{A}_z,\mbf{b}_z) \in \realsetmat{n}{n_g} \times \realset^n \times \realsetmat{n_c}{n_g} \times \realset^{n_c}$ such that
	\begin{equation} \label{eq:pre_cgrep}
		Z = \left\{ \mbf{c}_z + \mbf{G}_z \bm{\xi} : \| \bm{\xi} \|_\infty \leq 1, \mbf{A}_z \bm{\xi} = \mbf{b}_z \right\}.
	\end{equation}	
\end{definition}

In \eqref{eq:defhrep}, each inequality is a \emph{half-space}. In \eqref{eq:pre_cgrep}, each column of $\mbf{G}_z$ is a \emph{generator}, $\mbf{c}_z$ is the \emph{center}, and $\mbf{A}_z \bm{\xi} = \mbf{b}_z$ are the \emph{constraints}. We use compact notations $P = (\mbf{H}_p, \mbf{k}_p, \mbf{A}_p, \mbf{b}_p)\poly$ for H-rep polyhedra, $Z = (\mbf{G}_z, \mbf{c}_z,\mbf{A}_z,\mbf{b}_z)\czon$ for CZs, and $Z = (\mbf{G}_z, \mbf{c}_z)\zon$ for zonotopes. The latter two are referred to as the constrained generator representation (CG-rep) and generator representation (G-rep), respectively. Moreover, $(\mbf{H}_p, \mbf{k}_p, \noarg, \noarg)\poly$ and $(\noarg, \noarg, \mbf{A}_p, \mbf{b}_p)\poly$ denote polyhedra with only inequality constraints, and only equality constraints, respectively. An interval $X \in \intvalset^n$ can be described in G-rep as $(\text{diag}(\text{rad}(X)),\text{mid}(X))\zon$.

Consider sets $Z, W \subset \realset^{n}$, $Y \subset \realset^{m}$, and a matrix $\mbf{R} \in \realset^{m \times n}$. Define the Cartesian product, linear image, Minkowski sum, and generalized intersection, as $Z \times W \triangleq \{(\mbf{z},\mbf{w}): \mbf{z} \in Z, \mbf{w} \in W\}$, $\mbf{R}Z  \triangleq \{ \mbf{R} \mbf{z} : \mbf{z} \in Z\}$, $Z \oplus W  \triangleq \{ \mbf{z} + \mbf{w} : \mbf{z} \in Z,\, \mbf{w} \in W\}$, and $Z \cap_{\mbf{R}} Y  \triangleq \{ \mbf{z} \in Z : \mbf{R} \mbf{z} \in Y\}$, respectively. If $Z \triangleq (\mbf{G}_z, \mbf{c}_z, \mbf{A}_z, \mbf{b}_z)\czon \subset \realset^n$, $W \triangleq (\mbf{G}_w, \mbf{c}_w, \mbf{A}_w, $ $\mbf{b}_w)\czon \subset \realset^n$, and $Y \triangleq (\mbf{G}_y, \mbf{c}_y, \mbf{A}_y, \mbf{b}_y)\czon \subset \realset^m$ are constrained zonotopes, then 
\begin{align}
	Z {\times} W & = \! \left( \begin{bmatrix} \mbf{G}_z  & \bm{0} \\ \bm{0} & \mbf{G}_w \end{bmatrix}\!, \begin{bmatrix} \mbf{c}_z \\ \mbf{c}_w \end{bmatrix}\!, \begin{bmatrix} \mbf{A}_z & \bm{0} \\ \bm{0} & \mbf{A}_w \end{bmatrix}\!, \begin{bmatrix} \mbf{b}_z \\ \mbf{b}_w \end{bmatrix} \right)\czon\!\!\!\!, \label{eq:pre_czcartprod}\\
	\mbf{R}Z & = \left( \mbf{R} \mbf{G}_z, \mbf{R} \mbf{c}_z, \mbf{A}_z, \mbf{b}_z \right)\czon, \label{eq:pre_czlimage}\\
	Z {\oplus} W & = \left( [ \mbf{G}_z \,\; \mbf{G}_w ], \mbf{c}_z + \mbf{c}_w, \begin{bmatrix} \mbf{A}_z & \bm{0} \\ \bm{0} & \mbf{A}_w \end{bmatrix}\!, \begin{bmatrix} \mbf{b}_z \\ \mbf{b}_w \end{bmatrix} \right)\czon\!\!\!\!, \label{eq:pre_czmsum}\\
	Z {\cap_{\mbf{R}}} Y & = \left( [\mbf{G}_z \,\; \bm{0}], \mbf{c}_z, \begin{bmatrix} \mbf{A}_z & \bm{0} \\ \bm{0} & \mbf{A}_y \\ \mbf{R} \mbf{G}_z & -\mbf{G}_y \end{bmatrix}, \begin{bmatrix} \mbf{b}_z \\ \mbf{b}_y \\ \mbf{c}_y - \mbf{R} \mbf{c}_z \end{bmatrix} \right)\czon\!\!\!\!. \label{eq:pre_czintersection}
\end{align}

Moreover, define $B_\infty(\mbf{A}_z,\mbf{b}_z) \triangleq \{\bm{\xi} \in \realset^{n_g} : \ninf{\bm{\xi}} \leq 1,\,  \mbf{A}_z \bm{\xi} = \mbf{b}_z \}$. Then, $(\mbf{G}_z, \mbf{c}_z,\mbf{A}_z,\mbf{b}_z)\czon = \mbf{c}_z \oplus \mbf{G}_z B_\infty(\mbf{A}_z,\mbf{b}_z)$ holds, and $(\mbf{G}_z, \mbf{c}_z)\zon = \mbf{c}_z \oplus \mbf{G}_z B_\infty^n$. Additionally, if $P \triangleq (\mbf{H}_p, \mbf{k}_p, \mbf{A}_p, \mbf{b}_p)\poly \subset \realset^n$ and $Q \triangleq (\mbf{H}_q, \mbf{k}_q, \mbf{A}_q, \mbf{b}_q)\poly \subset \realset^n$, then
\begin{equation} \label{eq:hrepintersection}
	P \cap Q = \left( \begin{bmatrix} \mbf{H}_p \\ \mbf{H}_q \end{bmatrix}, \begin{bmatrix} \mbf{k}_p \\ \mbf{k}_q \end{bmatrix}, \begin{bmatrix} \mbf{A}_p \\ \mbf{A}_q \end{bmatrix}, \begin{bmatrix} \mbf{b}_p \\ \mbf{b}_q \end{bmatrix} \right)\poly.
\end{equation}

Efficient methods to enclose a CZ by another one with fewer generators and constraints are available in \cite{Scott2016}. The interval hull of a CZ, denoted as $\square Z \in \intvalset^n$, can be computed by solving $2n$ linear programs \citep{Scott2016,Rego2018}. Finally, the next result provides a method to obtain the exact intersection of a CZ and an H-rep polyhedron.

\begin{proposition} \rm \citep{Rego2024CZPRCDC} \label{prop:czhrepintersection}
	Let $Z = (\mbf{G}_z, \mbf{c}_z, \mbf{A}_z, \mbf{b}_z)\czon \subset \realset^n$ be a constrained zonotope and let $P \triangleq (\mbf{H}_p, \mbf{k}_p, \mbf{A}_p, \mbf{b}_p)\poly \subset \realset^n$ be a convex polyhedron in H-rep with $n_h$ halfspaces. Choose any $\bm{\sigma} \in \realset^{n_h}$ satisfying $\bm{\sigma} \leq \mbf{H}_p \mbf{z},~ \forall \mbf{z} \in Z$, and define $\mbf{G}_q \triangleq \half \text{diag} (\mbf{k}_p-\bm{\sigma})$, and $\mbf{c}_q \triangleq \half (\mbf{k}_p+\bm{\sigma})$. Then, 
	\begin{equation} \label{eq:czhrepintersection}
		\begin{aligned}
			Z \cap P = & \left( [\mbf{G}_z \,\; \mbf{0}], \mbf{c}_z, \begin{bmatrix} \mbf{A}_z & \mbf{0} \\ \mbf{H}_p \mbf{G}_z & - \mbf{G}_q  \\ \mbf{A}_p \mbf{G}_z & \mbf{0} \end{bmatrix}, \begin{bmatrix} \mbf{b}_z \\ \mbf{c}_q - \mbf{H}_p \mbf{c}_z \\ \mbf{b}_p - \mbf{A}_p \mbf{c}_z \end{bmatrix} \right)\czon.
		\end{aligned}
	\end{equation}
\end{proposition}

\begin{remark} \rm \label{rem:czhrepintersection}
 In this work, we calculate $\bm{\sigma}$ in Proposition \ref{prop:czhrepintersection} as the lower bound of the interval $\square(\mbf{H}_p\check{Z})$, where $\check{Z}$ is the zonotope obtained by neglecting the constraints of $Z$. This does not require the solution of linear programs since $\mbf{H}_p\check{Z}$ is a zonotope.
\end{remark}

\subsection{Problem formulation}

Consider the nonlinear discrete-time system
\begin{subequations} \label{eq:systemequations}
\begin{align}	
	\mbf{x}_k & = \mbf{f}(\mbf{x}_{k-1}, \mbf{w}_{k-1}, \mbf{u}_{k-1}), & & k\geq 1, \label{eq:systemf} \\
	\mbf{y}_k & = \mbf{g}(\mbf{x}_{k}, \mbf{v}_{k}), & & k\geq 0, \label{eq:systemg} 
\end{align}	
\end{subequations}
where $\mbf{x}_k \in \realset^{n_x}$ is the system state, $\mbf{u}_{k} \in \realset^{n_u}$ is a known input, $\mbf{y}_k \in \realset^{n_y}$ is the measured output, and $(\mbf{w}_k,\mbf{v}_k) \in \realset^{n_w+n_v}$ are bounded uncertainties. We assume that $\mbf{f}: \realset^{n_x} \times \realset^{n_w} \times \realset^{n_u} \to \realset^{n_x}$ and $\mbf{g}: \realset^{n_x} \times \realset^{n_v} \to \realset^{n_y}$ are factorable, and the initial condition and uncertainties are bounded by $(\mbf{x}_0,\mbf{w}_k,\mbf{v}_k) \in X_0 \times W \times V$, where $X_0$, $W$, and $V$ are known CZs.

The objective of this work is to obtain enclosures $\hat{X}_k$ of all states that are consistent with the initial set $X_0$, the uncertainty sets $W$ and $V$, the dynamics \eqref{eq:systemf}, and all measurements \eqref{eq:systemg} up to time $k$. For $k=0$, $\hat{X}_0$ must satisfy
\begin{equation} \label{eq:estimationk0}
\begin{aligned} 
	\hat{X}_0 \supseteq \{ \mbf{x}_0 \in X_0 : &~ \mbf{g}(\mbf{x}_0, \mbf{v}_0) = \mbf{y}_0,~ \mbf{v}_0 \in V\}.
\end{aligned}
\end{equation}
For $k \geq 1$, $\hat{X}_k$ must satisfy the recursion
\begin{equation} \label{eq:estimationrecursive}
\begin{aligned} 
	\hat{X}_k \supseteq \{ \mbf{x}_k \in \realset^{n_x} & : \mbf{x}_k = \mbf{f}(\mbf{x}_{k-1}, \mbf{w}_{k-1}, \mbf{u}_{k-1}), \ \mbf{y}_k = \mbf{g}(\mbf{x}_{k}, \mbf{v}_{k}),\\ & \qquad (\mbf{x}_{k-1}, \mbf{w}_{k-1},\mbf{v}_k) \in \hat{X}_{k-1} \times W \times V \}.
\end{aligned}
\end{equation}

With $\hat{X}_{k-1}$, $W$, and $V$ described as CZs, computing $\hat{X}_{k}$ satisfying \eqref{eq:estimationrecursive} requires propagating a CZ through the nonlinear function $\mbf{f}$ and then refining the resulting enclosure using the constraints $\mbf{g}(\mbf{x}_{k}, \mbf{v}_{k}) = \mbf{y}_k$.

\begin{remark} \rm \label{rem:invariants}
    The trajectories of real processes often satisfy solution invariants such as conservation of mass or energy. This information can often be written as equality constraints that can be used to refine $\hat{X}_{k}$ \citep{Yang2018EstManuf,Rego2021}. Invariants of the form $\mbf{h}(\mbf{x}_{k}, \mbf{d}_{k}) = \mbf{0}$ can be easily included in \eqref{eq:estimationk0}--\eqref{eq:estimationrecursive} provided that $\mbf{h}$ is factorable and $\mbf{d}_k$ is a bounded uncertainty. Such constraints can be included in \eqref{eq:systemg} as virtual measurement equations with measurement $\mbf{0}$.
\end{remark}

\section{Lifted polyhedral enclosure of factorable functions} \label{sec:liftedpolyhedron}

This section describes the core of the proposed state estimation method, which is a method for obtaining a polyhedral enclosure of the image of an interval under a nonlinear factorable function. This convex enclosure is later reformulated for CZ inputs, then refined with measurements and projected onto the state space. These additional operations require only intersections and affine transformations (see Section \ref{sec:stateestimation}).

Let $\bm{\varphi}: S \to \realset^{n_\varphi}$ be factorable with interval domain $S \in \intvalset^{n_s}$,  factors $\bm{\zeta}=(\zeta_1,\ldots,\zeta_{n_z})$, and $\mbf{E}_\varphi$ such that $\bm{\varphi}(\mbf{s}) = \mbf{E}_\varphi \bm{\zeta}(\mbf{s})$, $\forall\mbf{s} \in S$, as per Definition \ref{def:factorable}. We denote the image of $S$ under $\bm{\varphi}$ as $\bm{\varphi}(S)\triangleq\{\bm{\varphi}(\mbf{s}):\mbf{s}\in S\}$, and similarly for other sets and functions throughout. Note that $\bm{\varphi}(S)$ does not refer to the evaluation of $\bm{\varphi}$ in interval arithmetic. Our aim is to compute a convex polyhedron $P_\varphi \subset \realset^{n_z}$ enclosing $\bm{\zeta}(S)$. This provides an enclosure of $\bm{\varphi}(S)$ via
\begin{align}
\{\bm{\varphi}(\mbf{s}) : \mbf{s} \in S\} = \{\mbf{E}_\varphi \mbf{z}: \mbf{z} \in \bm{\zeta}(S)\} \subseteq \{\mbf{E}_\varphi \mbf{z}: \mbf{z} \in P_\varphi \}.
\end{align}

The following proposition shows that $P_\varphi$ can be calculated as the intersection of simple polyhedra $Q_j$ enclosing the graph of each elementary operation in the factorable representation of $\bm{\varphi}$. This result is similar to Theorem 4 in \cite{Siefert2025HZFuncDecomp}, which we refactor here due to notational differences.

\begin{proposition} \rm \label{thm:liftedpolyhedron}
    Let $\bm{\varphi}: S \to \realset^{n_\varphi}$ be a factorable function with interval domain $S \in \intvalset^{n_s}$ and factors $\bm{\zeta}=(\zeta_1,\ldots,\zeta_{n_z})$, and let $\bm{\alpha}=(\alpha_1,\ldots,\alpha_{n_z})$ be the corresponding elementary operations as in Definition \ref{def:factorable step functions}. Let $Z \in \intvalset^{n_z}$ satisfy $\bm{\zeta}(\mbf{s}) \in Z$, $\forall\mbf{s} \in S$. Moreover, for each $j>n_s$, let $Q_j \subset \realset^{n_z}$ satisfy
    \begin{align}
        Q_j \supseteq \{\mbf{z} \in Z : z_j = \alpha_{j}(\mbf{z})\}.
    \end{align}
    Then,
    \begin{equation} \label{eq:liftedpolyhedron}
    	\bm{\zeta}(S) \subseteq P_\varphi \triangleq \bigcap_{j=n_s+1}^{n_z} Q_j.
    \end{equation} 
    Moreover, $\bm{\varphi}(S)\subseteq \{\mbf{E}_\varphi \mbf{z}: \mbf{z} \in P_\varphi \}$.
\end{proposition}

\proof Choose any $\mbf{s} \in S$. By hypothesis, $\bm{\zeta}(\mbf{s}) \in Z$. Moreover, Definition \ref{def:factorable step functions} implies that $\zeta_j(\mbf{s}) = \alpha_{j}(\bm{\zeta}(\mbf{s}))$ for all $j \in \{1,\ldots,n_z\}$. It follows that $\bm{\zeta}(\mbf{s})\in Q_j$ for all $j \in \{n_s+1,\ldots,n_z\}$, and hence $\bm{\zeta}(\mbf{s})\in P_\varphi$. 
Since $\bm{\varphi}(\mbf{s}) = \mbf{E}_\varphi \bm{\zeta}(\mbf{s})$, we also have $\bm{\varphi}(\mbf{s}) \in \{\mbf{E}_\varphi \mbf{z}: \mbf{z} \in P_\varphi \}$. Since the choice of $\mbf{s}$ was arbitrary, $\bm{\zeta}(S) \subseteq P_\varphi$ and $\bm{\varphi}(S)\subseteq \{\mbf{E}_\varphi \mbf{z}: \mbf{z} \in P_\varphi \}$.
\qed

\begin{remark} \rm \label{rem:intervalZ}
In our implementation, the interval $Z$ required by Proposition \ref{thm:liftedpolyhedron} is computed as follows. For $j = 1,\ldots,n_s$, $Z_j \triangleq S_j \in \intvalset$. For $j>n_s$, except for certain cases discussed at the end of Section \ref{sec:nonlinearoperations}, the intervals $Z_j$ are obtained recursively using interval arithmetic. Specifically, either $Z_j = Z_a \odot Z_b$ or $Z_j=\beta_j(Z_a)$ according to Definition \ref{def:factorable step functions}.
\end{remark} 

The following subsections define simple polyhedral $Q_j$ for several common elementary operations. Once these are known, the intersection \eqref{eq:liftedpolyhedron} is computed trivially using \eqref{eq:hrepintersection}. 

\subsection{Arithmetic operations} \label{sec:arithmeticoperations}

Let $z_j = \alpha_j(\mbf{z}) = z_a\odot z_b$ with $a,b < j$ and $\odot\in\{+,-,\times,/\}$. Moreover, let $z_a \in Z_a \triangleq [\lb{z}_a, \ub{z}_a]$, and $z_b \in Z_b \triangleq [\lb{z}_b, \ub{z}_b]$.

\paragraph{Addition and subtraction}
$z_j = z_a \pm z_b$. In both cases, $Q_j$ is defined by the single equality constraint $z_j - z_a \pm z_b=0$. Specifically for addition, $Q_j = (\noarg,\noarg, \mbf{r}_+, 0)\poly$, where $\mbf{r}_+$ is the row vector such that $\mbf{r}_+ \mbf{z}=z_j-z_a - z_b$.

\paragraph{Multiplication}
$z_j = z_a z_b$. The H-rep enclosure $Q_j$ is obtained by rearranging the four inequalities in the McCormick convex envelope for $z_j = z_a z_b$, which are \citep{McCormick1976}
\begin{align*}
	\lb{z}_a z_b + z_a \lb{z}_b - \lb{z}_a \lb{z}_b &\leq z_j  \leq \lb{z}_a z_b + z_a \ub{z}_b - \lb{z}_a \ub{z}_b, \\
	\ub{z}_a z_b + z_a \ub{z}_b - \ub{z}_a \ub{z}_b &\leq z_j  \leq \ub{z}_a z_b + z_a \lb{z}_b - \ub{z}_a \lb{z}_b.
\end{align*}

\paragraph{Division}
$z_j = \frac{z_a}{z_b}$. The H-rep enclosure $Q_j$ is obtained by applying the multiplication rule above to $z_a = z_b z_j$. This requires an interval $Z_j\triangleq[\lb{z}_j,\ub{z}_j]$, which is computed by the interval division $Z_j = Z_a/Z_b$.

\paragraph{Addition or multiplication with a constant}
$z_j = z_a+q_j$ or $z_j = q_jz_a$, $q_j\in\mathbb{R}$. Strictly, these are univariate operations, but they are handled here for simplicity. In both cases $Q_j$ contains the single equality $z_j = z_a+q_j$ or $z_j = q_jz_a$.

\subsection{Univariate functions} \label{sec:nonlinearoperations}

Let $z_j = \alpha_j(\mbf{z}) = \beta_j(z_a)$, where $a < j$, $z_a \in Z_a \triangleq [\lb{z}_a, \ub{z}_a]$, and $\beta_j$ is a univariate function in the library $\mathcal{L}$. For all such functions, we assume that convex and concave relaxations on $Z_a$ can be readily constructed. Specifically, given any $Z_a\in\mathbb{IR}$, we have convex and concave functions $\beta_j^\text{CV} : Z_a \to \realset$ and $\beta_j^\text{CC} : Z_a \to \realset$, respectively, such that
\begin{equation}
	\label{Eq: Beta relaxations bound}
	\beta_j^\text{CV}(z_a) \leq \beta_j(z_a) \leq \beta_j^\text{CC}(z_a), \quad \forall z_a \in Z_a.
\end{equation}
Such relaxations are tabulated for a wide variety of common univariate functions in many global optimization references; see e.g.~Chapter 2 in \cite{ScottPhdThesis}.

Using these functions, we construct a polyhedral enclosure $Q_j = Q^\text{CV}_j \cap Q^\text{CC}_j$, where
\begin{alignat}{1}
	\label{Eq: QCV inclusion}
	Q^\text{CV}_j &\supseteq \{\mbf{z} \in \realset^{n_z} : z_j \geq \beta_j^\text{CV}(z_a), ~ z_a \in Z_a\}, \\
	\label{Eq: QCC inclusion}
	Q^\text{CC}_j &\supseteq \{\mbf{z} \in \realset^{n_z} : z_j \leq \beta_j^\text{CC}(z_a), ~ z_a \in Z_a\}.
\end{alignat}
Then, it holds that $Q_j \supseteq \{\mbf{z} \in \realset^{n_z} : z_j = \beta_{j}(z_a), ~ z_a \in Z_a\}$, as desired. Since $\beta_j^\text{CV}$ is convex and $\beta_j^\text{CC}$ is concave, the sets on the right-hand sides of \eqref{Eq: QCV inclusion}--\eqref{Eq: QCC inclusion} are enlarged if $\beta_j^\text{CV}$ and $\beta_j^\text{CC}$ are replaced by their linearizations at any point in $Z_a$. Therefore, our general strategy is to define $Q^\text{CV}_j$ and $Q^\text{CC}_j$ in terms of linearizations of $\beta_j^\text{CV}$ and $\beta_j^\text{CC}$ at a set of reference points. In many cases, these functions are linear, so no linearization is needed. Otherwise, we use linearizations at $\lb{z}_a$, $\midp{z}_a$, and $\ub{z}_a$. A few specific examples are given below.

\paragraph{Exponential}

$z_j = e^{z_a}$. In this case, $\beta_{j}(z_a)$ is convex, so $\beta^\text{CV}_{j}=\beta_{j}$. Thus, $Q^\text{CV}_j$ is obtained by linearizing $\beta_{j}$ at $\lb{z}_a$, $\midp{z}_a$, and $\ub{z}_a$, leading to the inequalities $z_j  \geq e^{\lb{z}_a} (z_a - \lb{z}_a) + e^{\lb{z}_a}$, $z_j  \geq e^{\midp{z}_a} (z_a - \midp{z}_a) + e^{\midp{z}_a}$, and $z_j  \geq e^{\ub{z}_a} (z_a - \ub{z}_a) + e^{\ub{z}_a}$. The concave relaxation on $Z_a$ is the secant $\beta_j^\text{CC}(z_a) = \left( \frac{e^{\ub{z}_a} - e^{\lb{z}_a}}{\ub{z}_a - \lb{z}_a} \right) (z_a - \lb{z}_a) + e^{\lb{z}_a}$. Thus, $Q^\text{CC}_j$ is defined by the single inequality $z_j  \leq \beta_j^\text{CC}(z_a)$.

\paragraph{Logarithm}

$z_j = \ln(z_a)$.  In this case, $\beta_{j}(z_a)$ is concave, so $\beta^\text{CC}_{j}=\beta_{j}$. Thus, $Q^\text{CC}_j$ is obtained by linearizing $\beta_{j}$ at $\lb{z}_a$, $\midp{z}_a$, and $\ub{z}_a$, leading to $z_j  \leq \frac{1}{\lb{z}_a} (z_a - \lb{z}_a) + \ln(\lb{z}_a)$, $z_j  \leq \frac{1}{\midp{z}_a} (z_a - \midp{z}_a) + \ln(\midp{z}_a)$, and $z_j  \leq \frac{1}{\ub{z}_a} (z_a - \ub{z}_a) + \ln(\ub{z}_a)$. The convex relaxation on $Z_a$ is the secant $\beta_j^\text{CV}(z_a) = \left( \frac{\ln(\ub{z}_a) - \ln(\lb{z}_a)}{\ub{z}_a - \lb{z}_a} \right) (z_a - \lb{z}_a) + \ln(\lb{z}_a)$.  Thus, $Q^\text{CV}_j$ is defined by the single inequality $z_j  \geq \beta_j^\text{CV}(z_a)$.

\paragraph{Even integer power}

$z_j = z_a^{q}$ with $q$ an even integer. In this case, $\beta_{j}(z_a)$ is convex, so $\beta^\text{CV}_{j}=\beta_{j}$. Thus, $Q^\text{CV}_j$ is obtained by linearizing $\beta_{j}$ at $\lb{z}_a$, $\midp{z}_a$, and $\ub{z}_a$, leading to the inequalities $z_j  \geq q(\lb{z}_a)^{(q-1)} (z_a - \lb{z}_a) + (\lb{z}_a)^q$, $z_j  \geq q(\midp{z}_a)^{(q-1)} (z_a - \midp{z}_a) + (\midp{z}_a)^q$, and $z_j  \geq q(\ub{z}_a)^{(q-1)} (z_a - \ub{z}_a) + (\ub{z}_a)^q$. The concave relaxation on $Z_a$ is the secant $\beta_j^\text{CC}(z_a) = \left( \frac{(\ub{z}_a)^q - (\lb{z}_a)^q}{\ub{z}_a - \lb{z}_a} \right) (z_a - \lb{z}_a) + (\lb{z}_a)^q$. Thus, $Q^\text{CC}_j$ is defined by the single inequality $z_j  \leq \beta_j^\text{CC}(z_a)$.

\paragraph{Odd integer power}

$z_j = z_a^{q}$ with $q$ an odd integer. In this case, $\beta_{j}(z_a)$ is concave for $z_a \leq 0$, and convex for $z_a \geq 0$. Therefore, for $\ub{z}_a \leq 0$, the enclosures $Q^\text{CV}_j$ and $Q^\text{CC}_j$ are obtained analogously to the logarithm, while for $\lb{z}_a \geq 0$, they are obtained analogously to the even integer power. If $0 \in [\lb{z}_a,\ub{z}_a]$, we use the following linearized versions of the piecewise relaxations in Chapter 2 of \cite{ScottPhdThesis}. Let $z^{\text{S}_1}_a$ be the solutions of $z_a^q - (\lb{z}_a)^q = (z_a - \lb{z}_a) q (z_a)^{q-1}$ with $z_a \in [0,\ub{z}_a]$. This is the point such that the slope of the secant between $\lb{z}_a$ and $z^{\text{S}_1}_a$ matches the slope of $z_a^q$ at $z^{\text{S}_1}_a$. Because $z_a^q$ is convex on $[0,\ub{z}_a]$, this secant underestimates $z_a^q$ to the right of $\lb{z}_a$, and hence on $Z_a$. If $\ub{z}_a\leq z^{\text{S}_1}_a$, then the secant from $\lb{z}_a$ to $\ub{z}_a$ is also a valid underestimator and is at least as tight. Thus, we define $\ub{\psi}_a \triangleq \min(\ub{z}_a,z^{\text{S}_1}_a)$ and enforce
\begin{equation}
    z_j \geq \left(\frac{(\ub{\psi}_a)^q - (\lb{z}_a)^q}{\ub{\psi}_a - \lb{z}_a}\right)(z_a - \lb{z}_a) + (\lb{z}_a)^q.
\end{equation}
If $\ub{z}_a>z^{\text{S}_1}_a$, then we add two more inequalities requiring $z_j$ to lie above the linearizations of $z_a^q$ at $\ub{z}_a$ and $\psi^\text{UM}_a \triangleq \text{mid}([z^{\text{S}_1}_a, \ub{z}_a])$. These (at most) three inequalities make up $Q_j^\text{CV}$. Analogously for $Q_j^\text{CC}$, let $z^{\text{S}_2}_a$ be the solution of $(\ub{z}_a)^q - z_a^q = (\ub{z}_a - z_a) q (z_a)^{q-1}$ with $z_a \in [\lb{z}_a,0]$. Letting $\lb{\psi}_a \triangleq \max(\lb{z}_a,z^{\text{S}_2}_a)$, we enforce $z_j \leq \left(\frac{(\ub{z}_a)^q - (\lb{\psi}_a)^q}{\ub{\psi}_a - \lb{z}_a}\right)(z_a - \lb{z}_a) + (\lb{z}_a)^q$. If $z^{\text{S}_2}_a > \lb{z}_a$, then we add two more inequalities requiring $z_j$ to lie below the linearizations of $z_a^q$ at $\lb{z}_a$ and $\psi^\text{LM}_a \triangleq \text{mid}([\lb{z}_a, z^{\text{S}_2}_a])$.

\paragraph{Sine}

$z_j = \sin(z_a)$. In this case, $\beta_{j}(z_a)$ is a periodic function alternating between convexity and concavity. For this reason, the interval $[\lb{z}_a, \ub{z}_a]$ is first partitioned into subregions according to the conditions found in Chapter 2 of \cite{ScottPhdThesis}, with valid convex and concave relaxations over the whole interval defined piecewise with different formulas for each subregion. Then, $Q^\text{CV}_j$ and $Q^\text{CC}_j$ are obtained by the intersection of multiple halfspaces, which require linearization of the formulas at some points in each subregion. Details are given in \ref{app:sine}.

\paragraph{Cosine} $z_j = \cos(z_a)$. Since $\cos(z_a) = \sin(z_a + \frac{\pi}{2})$, $Q_j$ is obtained by letting $z_b \triangleq z_a + \frac{\pi}{2}$ and $z_j = \sin(z_b)$ and using the methods for addition and sine.

\begin{figure}[!thb]
	\centering{
		\def\svgwidth{0.6\columnwidth}
  {\small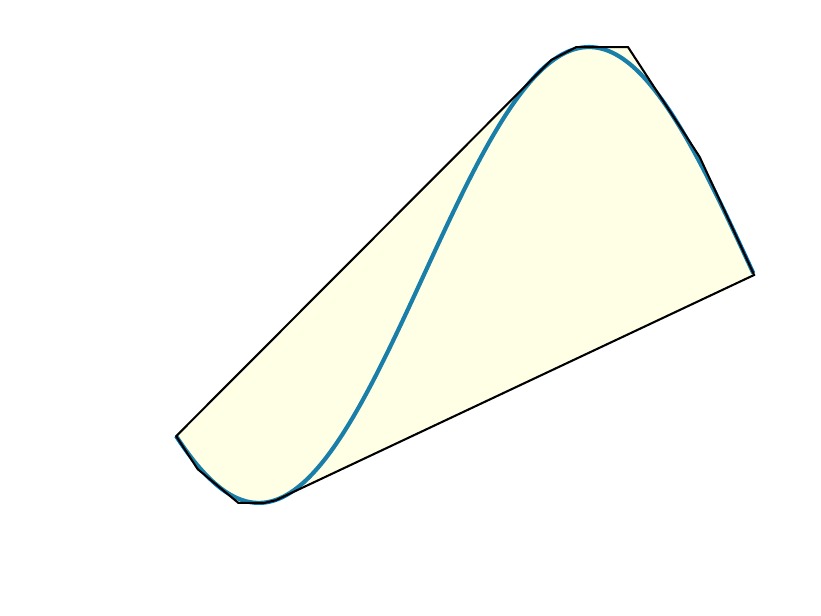}
		\caption{Halfspace enclosure $Q_j$ (yellow) of the set $\{(z_j,z_a) \in \realset \times Z_a : z_j = \sin(z_a)\}$ with $Z_a = [-\frac{3 \pi}{4}, \pi]$ (blue line).}\label{fig:liftedpolyhedronpow3}}
\end{figure}

Figure \ref{fig:liftedpolyhedronpow3} shows $Q_j$ for $z_j = \sin(z_a)$ with $z_a \in [-\frac{3 \pi}{4}, \pi]$. The convex hull of the graph is tightly approximated with limited complexity. Enabling the computation of polyhedral enclosures for trigonometric functions is an important contribution with respect to our previous work \citep{Rego2024CZPRCDC} because these functions often appear in vehicle and robotics applications.

The library $\mathcal{L}$ may include user-defined non-elementary univariate functions $\beta_j$ as long as methods for computing $Z_j$ and $Q_j$ are specified. Below we address three important cases for which polyhedral relaxations $Q_j$ can be built straightforwardly. We also discuss the computation of tighter intervals $Z_j$ for such cases, since the conservatism of each $Z_j$ can deteriorate the quality of the overall polyhedral relaxation.

\paragraph{Differentiable univariate convex function} $z_j = \beta_{j}(z_a)$, where $\beta_j$ is convex and differentiable in the interval $Z_a \triangleq [\lb{z}_a, \ub{z}_a]$. In this case, tight endpoints for the interval $Z_j \triangleq [\lb{z}_j, \ub{z}_j]$ can be obtained by minimum and maximum of $\beta_{j}$ over $z_a \in Z_a$. If any known local minimum\footnote{Either theoretically or obtained by guaranteed bisection search algorithms.} of $\beta_{j}$ belongs to $Z_a$, denoted by $z_a^\text{min}$, then $\lb{z}_j = \beta_{j}(z_a^\text{min})$, where $z_a^\text{min}$ is a point in $Z_a$ satisfying $\beta_j'(z_a^\text{min}) = 0$ (from convexity in $Z_a$), with $\beta_j' \triangleq \frac{d\beta_{j}}{dz_a}$. Otherwise, $\lb{z}_j = \text{min}\{\beta_j(\lb{z}_a),\beta_j(\ub{z}_a)\}$. In any case, $\ub{z}_j = \text{max}\{\beta_j(\lb{z}_a),\beta_j(\ub{z}_a)\}$. The convex relaxation $Q^\text{CV}_j$ on $Z_a$ is obtained by linearizing at $\beta_{j}$ at $\lb{z}_a$, $\midp{z}_a$, and $\ub{z}_a$. The concave relaxation $Q^\text{CC}_j$ on $Z_a$ is the secant leading to the inequality $z_j  \leq \beta_j^\text{CC}(z_a)$.

\paragraph{Differentiable univariate concave function} $z_j = \beta_{j}(z_a)$, where $\beta_j$ is concave and differentiable in the interval $Z_a \triangleq [\lb{z}_a, \ub{z}_a]$. The procedure is analogous to the previous case, in which the expressions for the lower and upper bounds (interval, convex and concave) are obtained by switching the former expressions from lower bounds to upper bounds and vice versa.%

\paragraph{Differentiable univariate DC function} $z_j = \beta_{j}(z_a)$, where $\beta_j$ is DC in $Z_a \triangleq [\lb{z}_a, \ub{z}_a]$. In other words, $\beta_j$ can be written as $\beta_j(z_a) = \psi_b(z_a) + \psi_c(z_a)$, where $\psi_b$ and $\psi_c$ are convex and concave in $Z_a$, respectively, and differentiable. Define intermediate variables $z_b = \psi_b(z_a)$ and $z_c = \psi_c(z_a)$. The intervals $z_b \in Z_b$ and $z_c \in Z_c$, along with the respective polyhedral relaxations, can be obtained using the aforementioned methods for convex and concave functions. The equality $z_j = z_b + z_c$ is included in the polyhedral relaxation for $z_j = \beta_{j}(z_a)$. In this special case, an interval $Z_j = [\lb{z}_j, \ub{z}_j]$ potentially tighter than the one computed by interval arithmetic is obtained using DC programming principles \citep{Alamo2008}. The endpoints $\lb{z}_j$ and $\ub{z}_j$ are given by optimizing the following expressions only over $\lb{z}_a$ and $\ub{z}_a$:
\begin{align}
    \lb{z}_j & = \underset{z \in \{\lb{z}_a, \ub{z}_a\}}{\text{min}} \psi_a(\midp{z}_a) + \psi_a'(\midp{z}_a)(z_a - \midp{z}_a) + \psi_b(z_a), \\
    \ub{z}_j & = \underset{z \in \{\lb{z}_a, \ub{z}_a\}}{\text{max}} \psi_a(z_a) + \psi_b(\midp{z}_a) + \psi_b'(\midp{z}_a)(z_a - \midp{z}_a).
\end{align}

\begin{figure}[!thb]
	\centering{
		\def\svgwidth{0.6\columnwidth}
  {\small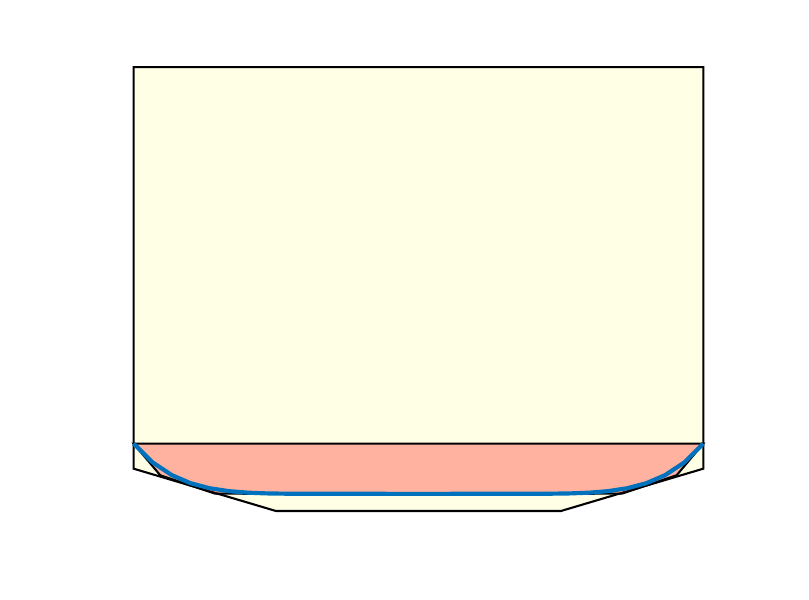}
		\caption{Halfspace enclosure $Q_j$ of the set $\{(z_j,z_a) \in \realset \times Z_a : z_j = (z_a^2 - z_a)^3\}$ with $Z_a = [-1,2]$ (blue line), obtained by decomposing each operation separately (yellow), and taking advantage that $z_a^2 - z_a$ is a convex function (orange).}\label{fig:liftedpolyhedronconvex}}
\end{figure}

To illustrate the advantages of considering these properties, Figure \ref{fig:liftedpolyhedronconvex} shows the improvements obtained by leveraging convexity to compute the polyhedral relaxation $Q_j$. 

\subsection{Complexity}
\label{sec:complexity of Pell}
Evaluating a function $\bm{\varphi}(\mbf{s})$ with $n_z$ factors requires evaluating $\tilde{n}_z=n_z-n_s$, elementary operations, and hence $\tilde{n}_z$ FLOPS (Floating-point Operations Per Second). Similarly, the cost of computing $Z$ and $P_{\varphi}$ is bounded by $c\tilde{n}_z$, where $c=c_Z+c_Q$ and $c_Z$ and $c_Q$ are the number of FLOPS required to compute a single $Z_j$ and $Q_j$, respectively. For arithmetic operations, $c_Z\leq8$ \citep{Moore2009} and $c_Q\leq 4$, with multiplication being the most costly. For all operations in Section \ref{sec:nonlinearoperations} except odd powers and trigonometrics, $c_Z\leq 6$ and $c_Q\leq 8$. Odd powers, $\sin$, and $\cos$ are not expensive, but may require iterative solutions of scalar equations, which makes $c_Q$ difficult to estimate. Overall, $c=10$ is a reasonable estimate. For comparison, computing a single element of the interval Jacobian of $\bm{\varphi}$ using forward automatic differentiation and interval arithmetic costs $\sim 10\tilde{n}_z$. This is done, e.g., in linearization methods using the mean value error bound \citep{Rego2021}.

The total number of constraints in $P_{\varphi}$ can similarly be estimated by $d\tilde{n}_z$, where $d$ is the number of constraints in a single $Q_j$. By inspection, $d=1$ for addition and subtraction, $d=4$ for multiplication, division, $\ln$, $\exp$, and even powers, $d\leq 6$ for odd powers, and $d\leq 19$ for $\sin$ and $\cos$. A reasonable average estimate is $d=10$. Note that the constraint matrices $\mbf{H}_{p_\varphi}$ and $\mbf{A}_{p_\varphi}$ describing $P_{\varphi}$ are sparse by construction, which helps mitigate the memory burden in high-dimensional applications.

\section{State estimation} \label{sec:stateestimation}

This section presents a new method, referred to as CZPR, for computing CZ enclosures $\hat{X}_k$ satisfying \eqref{eq:estimationrecursive} using the polyhedral relaxations from Section \ref{sec:liftedpolyhedron}. This method makes use of the computational advantages of constrained zonotopes in affine transformations and complexity reduction, while leveraging the efficiency with which polyhedral relaxations are obtained for factorable functions. The new methodology avoids the need of conservative approaches such as Taylor expansions.

Henceforth, let $\hat{X}_{k-1}$ be a known CZ enclosure at time $k-1$ and let $\mbf{y}_k$ be the measured output at time $k$. The goal is to compute an enclosure $\hat{X}_k$ of 
\begin{equation} \label{eq:estexactX_1}
\underline{X}_k \triangleq \left\{ \mbf{x}_k: 
\begin{aligned} 
    & \mbf{x}_k = \mbf{f}(\mbf{x}_{k-1}, \mbf{w}_{k-1}, \mbf{u}_{k-1}), \\ & \mbf{y}_k = \mbf{g}(\mbf{x}_{k}, \mbf{v}_{k}), \\ & (\mbf{x}_{k-1}, \mbf{w}_{k-1},\mbf{v}_k) \in \hat{X}_{k-1} \times W \times V 
\end{aligned}
\right\}.
\end{equation} 
Common practice in set-based state estimation is to handle this in two steps: First, the image of $\hat{X}_{k-1} \times W$ under $\mbf{f}(\cdot,\cdot, \mbf{u}_{k-1})$ is enclosed without considering the measurement equations; Second, the resulting enclosure is refined using the measurements. These are referred to as the prediction and update steps, respectively. Here, we combine these into a single \emph{prediction-update step} to streamline the presentation and implementation.

Define the composite function $\bm{\ell}: \realset^{n_x} \times \realset^{n_w} \times \realset^{n_v} \to \realset^{n_y}$ as $\bm{\ell}(\mbf{x}_{k-1},\mbf{w}_{k-1},\mbf{v}_k) \triangleq \mbf{g}(\mbf{f}(\mbf{x}_{k-1},\mbf{w}_{k-1},\mbf{u}_{k-1}),\mbf{v}_k)$. Since the control input $\mbf{u}_{k-1}$ is assumed to be known, it can be treated as a constant parameter in everything that follows and is therefore not included in the arguments to $\bm{\ell}$. For the remaining arguments, define the shorthand notations $\mbf{s} \triangleq (\mbf{x}_{k-1},\mbf{w}_{k-1},\mbf{v}_k)$ and $S \triangleq \hat{X}_{k-1} \times W \times V$. Since both $\mbf{f}$ and $\mbf{g}$ are factorable, $\bm{\ell}$ is also factorable. Let $\bm {\zeta} = (\zeta_1, \ldots, \zeta_{n_z})$ be the factors of $\bm{\ell}$ and let $\mbf{E}_\ell$ be the matrix from Definition \ref{def:factorable} such that $\bm{\ell}(\mbf{s}) = \mbf{E}_\ell\bm{\zeta}(\mbf{s})$,  $\forall \mbf{s} \in S$. By the construction of $\bm{\ell}$, $\mbf{f}$ can also be described in terms of the factors $\bm{\zeta}$. As such, there also exists $\mbf{E}_f$ such that $\mbf{f}(\mbf{s},\mbf{u}_{k-1}) = \mbf{E}_f\bm{\zeta}(\mbf{s})$.
Using these definitions, the set we wish to enclose can be written more simply as
\begin{alignat}{1}
    \underline{X}_k &= \{\mbf{f}(\mbf{s},\mbf{u}_{k-1}) : \bm{\ell}(\mbf{s}) = \mbf{y}_k, \ \mbf{s} \in S\}, \label{eq:estexactX_2} \\
    &= \{\mbf{E}_f\bm{\zeta}(\mbf{s}) : \mbf{E}_{\ell}\bm{\zeta}(\mbf{s}) = \mbf{y}_k, \ \mbf{s} \in S\}. \label{eq:estexactX_3}
\end{alignat}

Let $\Box S$ denote the interval hull of $S$ and recall that an interval $Z$ enclosing the image $\bm{\zeta}(\Box S)$ can be readily obtained from interval arithmetic as discussed in Remark \ref{rem:intervalZ}. Moreover, Proposition \ref{thm:liftedpolyhedron} provides a method for using $Z$ to obtain a tighter polyhedral enclosure $\bm{\zeta}(\Box S)\subseteq P_{\ell}$. The next proposition shows how $Z$ and $P_{\ell}$ can be used to form an enclosure of $\underline{X}_k$. While using different enclosures and providing different results, this proposition carries a few similarities with Theorem 2 in \cite{SiefertHZEstimation2023} and Corollary 2 in \cite{Siefert2025HZFuncDecomp}. The numerical calculation of this enclosure is discussed subsequently.

\begin{proposition} \rm \label{thm:predictionupdate}
Let $Z\in\intvalset^{n_z}$ and $P_{\ell}\subset\realset^{n_z}$ be interval and polyhedral enclosures of $\bm{\zeta}(\Box S)$, respectively. Let $\tilde{Z}$ denote the last $n_z - (n_x+n_w+n_v)$ components of the interval vector $Z$. Finally, define $P_y \triangleq (\noarg, \noarg, \mbf{E}_\ell, \mbf{y}_k)\poly$. Then,
\begin{equation} \label{eq:czprpredictionupdate}
\underline{X}_k \subseteq\hat{X}_k \triangleq \mbf{E}_f ((\hat{X}_{k-1} \times W \times V \times \tilde{Z}) \cap P_\ell \cap P_y ).
\end{equation}
\end{proposition}

\proof Choose any $\mbf{x}_k\in \underline{X}_k$. By \eqref{eq:estexactX_3}, $\exists\mbf{s}\in S$ such that
\begin{alignat}{1}
\mbf{x}_k &= \mbf{E}_f\bm{\zeta}(\mbf{s}), \label{eq: state estimator proof - xk=f(s)} \\
\mbf{y}_k &= \mbf{E}_\ell\bm{\zeta}(\mbf{s}). \label{eq: state estimator proof - yk=ell(s)}
\end{alignat}
We now show that $\bm{\zeta}(\mbf{s})$ is contained in all three sets that are intersected on the right-hand side of \eqref{eq:czprpredictionupdate}. Equation \eqref{eq: state estimator proof - yk=ell(s)} immediately implies that $\bm{\zeta}(\mbf{s})\in P_{y}$. Since $\mbf{s}\in S\subseteq\Box S$, we also have $\bm{\zeta}(\mbf{s})\in P_{\ell}$ by the hypothesis on $P_{\ell}$. Now consider the set $\hat{X}_{k-1} \times W \times V \times \tilde{Z}$, which is the same as $S\times \tilde{Z}$. The last $n_z-(n_x+n_w+n_v)$ components of  $\bm{\zeta}(\mbf{s})$ lie in $\tilde{Z}$ by hypothesis. Then, since $S\times \tilde{Z}$ is a Cartesian product, it suffices to show that the first $(n_x+n_w+n_v)$ components of $\bm{\zeta}(\mbf{s})$ lie in $S$. But, by Definition \ref{def:factorable}, the first $(n_x+n_w+n_v)$ components of $\bm{\zeta}(\mbf{s})$ are exactly $\mbf{s}\in S$. Thus, $\bm{\zeta}(\mbf{s})$ is contained in $(S \times \tilde{Z}) \cap P_\ell \cap P_y$. By \eqref{eq:czprpredictionupdate}, this implies that $\mbf{E}_f\bm{\zeta}(\mbf{s})\in \hat{X}_k$. By \eqref{eq: state estimator proof - xk=f(s)}, it follows that $\mbf{x}_k\in \hat{X}_k$. Since $\mbf{x}_k\in \underline{X}_k$ was chosen arbitrarily, $\underline{X}_k\subseteq \hat{X}_k$. \qed

The enclosure in \eqref{eq:czprpredictionupdate} can be computed efficiently as follows. First, $\Box S$ is computed by solving $2n_s$ LPs, where $n_s=n_x+n_w+n_v$. Assume that $\hat{X}_{k-1}$ has $n_{g_{k-1}}$ generators and $n_{c_{k-1}}$ constraints, so that $S$ has $n_{g_s}=n_{g_{k-1}} + n_{g_w} + n_{g_v}$ and $n_{c_s}=n_{c_{k-1}} + n_{c_w} + n_{c_v}$. The complexity of computing $\Box S$ is polynomial in $n_s$, $n_{g_s}$, and $n_{c_s}$. A cheaper and more conservative interval enclosure could clearly be used, but we do not do so here. Next, $Z$ and $P_{\ell}$ are computed as in Remark \ref{rem:intervalZ} and Proposition \ref{thm:liftedpolyhedron} at a cost of $O(\tilde{n}_z)$, where $\tilde{n}_z = n_z - n_s$ (see Sec.~\ref{sec:complexity of Pell}). The intersection $P_{\ell}\cap P_y$ is then computed by \eqref{eq:czhrepintersection}, and the Cartesian product $\hat{X}_{k-1} \times W \times V \times \tilde{Z}$ is easily computed by \eqref{eq:pre_czcartprod}. These two steps require matrix concatenations, but no computations. The intersection of $\hat{X}_{k-1} \times W \times V \times \tilde{Z}$ with ($P_{\ell}\cap P_y$) is computed using \eqref{eq:czhrepintersection}. Assume that $P_\ell \cap P_y$ has $n_{h_p}$ halfspaces and $n_{c_p}$ equality constraints ($P_y$ contributes $n_y$ equalities and no inequalities). Then, this step has complexity $O((n_{h_p}+n_{c_p})n_z(n_{g_s} + \tilde{n}_z))$, or simply $O(n_z^2(n_{g_s} + \tilde{n}_z))$ because $(n_{h_p}+n_{c_p})\propto \tilde{n}_z$ (see Sec.~\ref{sec:complexity of Pell}). The resulting CZ is finally mapped under $\mbf{E}_f$ to obtain $\hat{X}_k$. Since each row of $\mbf{E}_f$ is a unit vector, this step requires no computations.

The numbers of generators and constraints in $\hat{X}_k$ are
\begin{alignat}{1}
n_{g_k} &= n_{g_{k-1}} + n_{g_w} + n_{g_v} + \tilde{n}_z + n_{h_p},  \label{eq: predUpdate set complexity - ng } \\
n_{c_k} &= n_{c_{k-1}} + n_{c_w} + n_{c_v} + n_{h_p} + n_{c_p}. \label{eq: predUpdate set complexity - nc }
\end{alignat}
Notably, the values $\tilde{n}_z$, $n_{h_p}$, and $n_{c_p}$ are constant and depend only on the (non-unique) factorization of $\bm{\ell}$. This implies that the complexity increase at each time step is linear, which is an important advantage compared to some other methods. For example, CZ methods that use linearization with error bounds from Taylor's Theorem have quadratic complexity growth \citep{Rego2021}. Worse, methods that directly propagate H-rep polyhedra have exponential complexity growth. Typically, all growth is eliminated by applying complexity reduction methods in each time step \citep{Scott2016}. Thus, the advantage of slower growth is that it requires less reduction, which avoids additional conservatism and reduces computational cost. The complexities to eliminate $k_g$ generators and $k_c$ constraints from $\hat{X}_k$ are $O((n_x+n_{c_k})^2 n_{g_k} + k_g {n_{g_k}}(n_x+n_{c_k}))$ and $O(k_c(n_{g_k} + n_{c_k})^3)$, respectively. Since $k_g = n_{g_k} - n_{g_{k-1}}$ and $k_c = n_{c_k} - n_{c_{k-1}}$ in steady-state, if $n_{g_k}$ and $n_{c_k}$ are given by \eqref{eq: predUpdate set complexity - ng }--\eqref{eq: predUpdate set complexity - nc }, then order reduction has third-order polynomial complexity in $(n_{g_{k-1}},n_{c_{k-1}},k_g)$, and order four in $k_c$ (this constitutes the algorithmic step with the highest complexity in our method). A quadratic growth would mean $n_{g_k} \propto n_{g_{k-1}}^2$, $n_{c_k} \propto n_{c_{k-1}}^2$, implying $k_g \propto n_{g_{k-1}}^2$, $k_c \propto n_{c_{k-1}}^2$, resulting in order eight, for instance.

\begin{remark} \rm \label{rem:use of interval to compute P}
The proposed approach for constructing $P_\ell$ requires an interval domain, which forces us to use an interval enclosure of $\hat{X}_{k-1}\times W\times V$. Thus, any previous dependencies between states is lost in the computation of $P_\ell$. However, this information is later used directly in the intersection with $\hat{X}_{k-1}\times W\times V$ in \eqref{eq:czprpredictionupdate}. This is somewhat analogous to the situation with existing set-based estimation methods based on conservative linearization, where one is forced to use an interval enclosure of $\hat{X}_{k-1}\times W\times V$ to bound the Jacobian or Hessian in the course of bounding the linearization error \citep{Rego2021}.
\end{remark}

\begin{remark} \rm \label{rem:separatesteps}
In some applications, it may be preferable to compute the prediction and update steps separately, as done in previous literature \citep{Alamo2005a,Rego2021}. This can be done by straightforward modifications of Proposition \ref{thm:predictionupdate}. To enclose the prediction set $\underline{X}_k^p\triangleq\{\mbf{f}(\mbf{x}_{k-1}, \mbf{w}_{k-1}, \mbf{u}_{k-1}) : (\mbf{x}_{k-1}, \mbf{w}_{k-1}) \in \hat{X}_{k-1} \times W\}$, let $\bm{\delta} = (\delta_1,\ldots,\delta_{n_\delta})$ be the factors of $\mbf{f}(\cdot,\cdot,\mbf{u}_{k-1})$, let $\Delta$ and $P_{f}$ be interval and polyhedral enclosures of the image $\bm{\delta}(\Box (\hat{X}_{k-1} \times W))$ obtained as in Remark \ref{rem:intervalZ} and Proposition \ref{thm:liftedpolyhedron}, and let $\tilde{\Delta}$ denote the last $n_\delta - (n_x+n_w)$ elements of $\Delta$. Then, $\underline{X}_k^p \subseteq \hat{X}_k^p \triangleq \mbf{E}_f ((\hat{X}_{k-1} \times W \times \tilde{\Delta}) \cap P_f)$. To enclose the update set $\underline{X}_k^u\triangleq\{\mbf{x}_k \in \hat{X}_k^p : \mbf{y}_k = \mbf{g}(\mbf{x}_k, \mbf{v}_k)$, $\mbf{v}_k \in V \}$, define $\mbf{s} \triangleq (\mbf{x}_k,\mbf{v}_k)$ and $S \triangleq \hat{X}_k^p\times V$. Let $\bm{\gamma} = (\gamma_1,\ldots,\gamma_{n_\gamma})$ be the factors of $\mbf{g}$ so that $\mbf{g}(\mbf{s}) = \mbf{E}_g \bm{\gamma}(\mbf{s})$, $\forall \mbf{s}\in S$. Let $\Gamma$ and $P_{g}$ be interval and polyhedral enclosures of the image $\bm{\gamma}(\Box S)$ obtained as in Remark \ref{rem:intervalZ} and Proposition \ref{thm:liftedpolyhedron}, and let $\tilde{\Gamma}$ denote the last $n_\gamma - (n_x+n_v)$ elements of $\Gamma$. Finally, define $P_y \triangleq (\noarg, \noarg, \mbf{E}_g, \mbf{y}_k)\poly$. Then, $\underline{X}_k^u\subseteq \hat{X}_k \triangleq \mbf{E}_x ((\hat{X}_k \times V \times \tilde{\Gamma}) \cap (P_g \cap P_y ))$, where $\mbf{E}_x \triangleq [\eye{n_x} \,\; \zeros{n_x}{(n_\gamma - n_x)}]$. Note, however, that this approach computes the interval hull of a CZ twice per $k$ instead of once (once for $\hat{X}_k\times W$ and again for $\hat{X}_k^p\times V$), which is undesirable because it requires the solution of linear programs. On the other side, in control applications where execution times are critical, the prediction set $\hat{X}_k^p$, which requires only measurements up to $k-1$, can be used to generate $\mbf{u}_k$. However, this is a weaker enclosure and potentially leads to more conservative control.
\end{remark}

\begin{remark} \rm \label{rem:czprinitialupdate}
    An enclosure of the initial set $\underline{X}_0\triangleq\{\mbf{x}_0 \in X_0 : \mbf{y}_0 = \mbf{g}(\mbf{x}_0, \mbf{v}_0)$, $\mbf{v}_0 \in V \}$ can be obtained as a special case of the update step in Remark \ref{rem:separatesteps}.
\end{remark}

\begin{remark} \rm \label{rem:linearupdate}
When the measurement equations are linear, i.e., $\mbf{g}(\mbf{x}_k,\mbf{v}_k) = \mbf{C} \mbf{x}_k + \mbf{D}_v \mbf{v}_k$, Proposition \ref{thm:predictionupdate} leads to some unnecessary computations. A better approach is to compute a prediction enclosure $\hat{X}_k^p$ following Remark \ref{rem:separatesteps} and then execute the linear update $\hat{X}_k = \hat{X}_k^p \cap_\mbf{C}(\mbf{y}_k \oplus (-\mbf{D}_vV))$ using standard CZ operations.
\end{remark}

\subsection{Equivalent representation of $\hat{X}_k$ with reduced complexity} \label{sec:equivalentset}

Although the complexity of $\hat{X}_k$ grows only linearly with $k$ under Proposition \ref{thm:predictionupdate}, it can nevertheless become prohibitive when $n_z$ is large (i.e., when $\mbf{f}$ or $\mbf{g}$ involve a large number of elementary operations). Since each operation adds constraints to $P_{\ell}$, this makes $P_{\ell}$ complex and ultimately contributes many generators and constraints to $\hat{X}_k$. To mitigate this, this section describes a method for automatically reducing the number of equality constraints in $P_\ell \cap P_y$ with no added conservatism. The procedure also reduces the dimension of $P_\ell \cap P_y$ and $Z$. To improve the reader guidance, the necessary proofs are omitted (these can be found in \ref{app:proofs}).

Let $\mbf{s}$, $S$, $\bm{\ell}$, $\bm{\zeta}$, $\mbf{E}_{\ell}$, and $\mbf{E}_f$ be defined as in Section \ref{sec:stateestimation}. Furthermore, define $Z$, $\tilde{Z}$, $P_{\ell}$, and $P_y$ as in Proposition \ref{thm:predictionupdate}. The results in this section require the following assumption.

\begin{assumption} \rm \label{ass:intervalZ}
    The interval $Z$ is computed by applying interval arithmetic to the factorable representation of $\bm{\ell}$ as described in Remark \ref{rem:intervalZ}. Moreover, $P_{\ell}$ is computed as in Proposition \ref{thm:liftedpolyhedron}. Specifically, the $Q_j$ used for addition, subtraction, addition of a constant, and multiplication by a constant are all exactly as given in Section \ref{sec:arithmeticoperations}.
\end{assumption}

Let $\bm{\alpha}=(\alpha_1,\ldots,\alpha_{n_z})$ be the elementary operations defining $\bm{\ell}$ (see Definition \ref{def:factorable step functions}). Let $\mathcal{I} \triangleq \{1,2,\ldots,n_z\}$ and define $\mathcal{I}_\text{e}$ as the subset of indices in $\mathcal{I}$ such that $\alpha_j$ is given by one of the
following operations, where $q_j \in \realset$ is a constant:
\begin{equation} \label{eq:operationstobesolved}
\begin{aligned}
& \alpha_j(z_a,z_b) = z_a + z_b, & & \alpha_j(z_a,z_b) = z_a + q_j, \\
& \alpha_j(z_a,z_b) = z_a - z_b, & &
\alpha_j(z_a,z_b) = q_j z_a.
\end{aligned}
\end{equation}
The significance of these operations is that they all contribute a single equality constraint to $P_{\ell}$ (and hence to $P_{\ell}\cap P_y$) that specifies $z_j$ exactly as a linear function of $(z_a,z_b)$. Our aim is to eliminate these constraints, and the corresponding $z_j$, from the representation of $P_{\ell}\cap P_y$ by substituting these relations in wherever $z_j$ appears in the other constraints defining $P_{\ell}\cap P_y$. We say that these $z_j$ are \emph{eliminated} (hence the `e' in $\mathcal{I}_\text{e}$), while the rest are \emph{retained}. Define $\mathcal{I}_\text{r}\triangleq\mathcal{I} \backslash \mathcal{I}_\text{e}$.

Let $(\mbf{H}_p, \mbf{k}_p, \mbf{A}_p, \mbf{b}_p)\poly$ denote the extended H-rep of $P_{\ell}\cap P_y$. Let $\mbf{U}\in\mathbb{R}^{n_z\times n_z}$ be the permutation matrix that rearranges any $\mbf{z}\in\mathbb{R}^{n_z}$ as $\mbf{Uz}=(\mbf{z}_\text{e},\mbf{z}_\text{r})$, where $\mbf{z}_\text{e} \in \realset^{n_\text{e}}$ contains all $z_j$ with $j \in \mathcal{I}_\text{e}$, in ascending order, and $\mbf{z}_\text{r} \in \realset^{n_\text{r}}$ contains all $z_j$ with $j \in \mathcal{I}_\text{r}$, also in ascending order. Similarly, let $\mbf{V}\in\mathbb{R}^{n_{c_p}\times n_{c_p}}$ be the permutation such that $\mbf{V}\mbf{A}_p=\left[\begin{smallmatrix}\mbf{A}_\text{e} \\ \mbf{A}_\text{r}\end{smallmatrix}\right]$, where $\mbf{A}_\text{e}$ contains the rows of $\mbf{A}_p$ corresponding to $Q_j$ for each $j \in \mathcal{I}_\text{e}$, in ascending order, and $\mbf{A}_\text{r}$ contains the remaining rows with their original ordering. Note that each $Q_j$ with $j\in\mathcal{I}_\text{e}$ contains a single equality constraint, so $\mbf{A}_\text{e}$ has exactly $n_\text{e}=|\mathcal{I}_\text{e}|$ rows. With these definitions, $\mathbf{A}_p\mbf{z}=\mbf{b}_p$ is equivalent to $(\mbf{VA}_p\mbf{U}^{-1})(\mbf{Uz})=\mbf{Vb}_p$, which can be written in terms of transformed variables as 
\begin{equation} \label{eq:equalitiespartition}
\begin{bmatrix} \mbf{A}_\text{ee} & \mbf{A}_\text{er} \\ \mbf{A}_\text{re} & \mbf{A}_\text{rr} \end{bmatrix} \begin{bmatrix} \mbf{z}_\text{e} \\ \mbf{z}_\text{r} \end{bmatrix} = \begin{bmatrix} \mbf{b}_\text{e} \\ \mbf{b}_\text{r} \end{bmatrix}.
\end{equation}
Note that $\mbf{U}^{-1}$ permutes the columns of $\mbf{A}_p$ to match the permutation of $\mbf{z}$. Thus, $\mathbf{A}_\text{ee}$ contains the columns of $\mathbf{A}_\text{e}$ indexed by $j\in\mathcal{I}_\text{e}$ in ascending order. Since $\mathbf{A}_\text{e}$ has $n_\text{e}$ rows, $\mathbf{A}_\text{ee}$ is square.

\begin{lemma} \rm
\label{lem:Aee properties}
$\mathbf{A}_\text{ee}$ is lower triangular with ones on its diagonal.
\end{lemma}
\proof See \ref{app:proofs}. \qed

An important consequence of Lemma \ref{lem:Aee properties} is that $\mbf{A}_\text{ee}$ is invertible. Thus, the first block row of \eqref{eq:equalitiespartition} can be solved for $\mbf{z}_{\text{e}}$ as
\begin{equation} \label{eq:factorelimination}
    \mbf{z}_\text{e} = \mbf{A}_\text{ee}^{-1} (\mbf{b}_\text{e} - \mbf{A}_\text{er} \mbf{z}_\text{r}).
\end{equation}
The next lemma shows that this can be used to obtain an equivalent description of $P_{\ell}\cap P_y$ without these variables and constraints. In preparation for this result, define $[\mbf{H}_\text{e} \  \mbf{H}_\text{r}]=\mbf{H}_p\mbf{U}^{-1}$ and note that the inequalities $\mbf{H}_p \mbf{z} \leq \mbf{k}_p$ can also be written equivalently in transformed variables as
\begin{equation}
\label{eq:inequalitiespartition}
    \mbf{H}_\text{e} \mbf{z}_\text{e} + \mbf{H}_\text{r} \mbf{z}_\text{r} \leq \mbf{k}_p.
\end{equation}

\begin{lemma} \rm \label{thm:equivalentpolytope}
Define the polyhedron $\mathring{P} \triangleq (\mathring{\mbf{H}},\mathring{\mbf{k}},\mathring{\mbf{A}},\mathring{\mbf{b}})\poly$, where
\begin{equation} \label{eq:equivalentpolytope}
\begin{aligned}
    & \mathring{\mbf{H}} \triangleq (\mbf{H}_\text{r} - \mbf{H}_\text{e} \mbf{A}_\text{ee}^{-1}  \mbf{A}_\text{er}), & & \mathring{\mbf{k}} \triangleq \mbf{k}_p - \mbf{H}_\text{e} \mbf{A}_\text{ee}^{-1}  \mbf{b}_\text{e}, \\
    & \mathring{\mbf{A}} \triangleq (\mbf{A}_\text{rr} - \mbf{A}_\text{re} \mbf{A}_\text{ee}^{-1}  \mbf{A}_\text{er}), & & \mathring{\mbf{b}} \triangleq \mbf{b}_\text{r} - \mbf{A}_\text{re} \mbf{A}_\text{ee}^{-1}  \mbf{b}_\text{e}. 
\end{aligned}
\end{equation}
For any $\mbf{z}\in\realset^{n_z}$ with $(\mbf{z}_\text{e},\mbf{z}_\text{r})=\mbf{Uz}$,
\begin{equation}
    \mbf{z} \in P_\ell \cap P_y \iff 
    \mbf{z}_\text{r} \in \mathring{P} \text{ and }
    \eqref{eq:factorelimination}\text{ holds}.
\end{equation}
\end{lemma}
\proof See \ref{app:proofs}. \qed

To convert this result into a simpler enclosure for $\underline{X}_k$, let $(Z_\text{e},Z_\text{r}) = \mbf{U}Z$. Note that this multiplication merely reorders the elements of $Z$ and does not require any interval arithmetic. %
Moreover, define $[\mbf{E}_\text{e} \  \mbf{E}_\text{r}]=\mbf{E}_f\mbf{U}^{-1}$ so that, for all $\mbf{z}\in \mathbb{R}^{n_z}$,
\begin{alignat}{1}
    \label{eq:Efpartition}
    \mbf{E}_f\mbf{z} &= (\mbf{E}_f\mbf{U}^{-1})(\mbf{U}\mbf{z}) = \mbf{E}_\text{e} \mbf{z}_\text{e} + \mbf{E}_\text{r} \mbf{z}_\text{e}.
\end{alignat}

\begin{theorem} \rm
\label{lem:newpredictionupdate}
Let $\mathring{P}$ be defined as in Lemma \ref{thm:equivalentpolytope} and let $\tilde{Z}_\text{r}$ denote the last $n_r - (n_x+n_w+n_v)$ elements of $Z_\text{r}$. Define
\begin{equation} \label{eq:czprnewpredictionupdate}
    \hat{X}_k' \triangleq \mbf{G}_f ((\hat{X}_{k-1} \times W \times V \times \tilde{Z}_\text{r}) \cap \mathring{P}) \oplus \mbf{c}_f,
\end{equation}
where
\begin{equation} \label{eq:gfcf}
    \mbf{G}_f \triangleq (\mbf{E}_\text{r} - \mbf{E}_\text{e} \mbf{A}_\text{ee}^{-1}  \mbf{A}_\text{er}), \quad \mbf{c}_f \triangleq (\mbf{E}_\text{e} \mbf{A}_\text{ee}^{-1}  \mbf{b}_\text{e}).
\end{equation}
Then: 
\begin{enumerate}
\item $\hat{X}_k'$ is a valid enclosure of $\underline{X}_k$.
\item Let $\hat{X}_k$ be the enclosure defined in Proposition \ref{thm:predictionupdate}. If $\mbf{A}_\text{ee}^{-1} (\mbf{b}_\text{e} - \mbf{A}_\text{er} \mbf{z}_\text{r})\in Z_\text{e}$ for all $\mbf{z}_\text{r}\in Z_\text{r}$, then $\hat{X}_k'=\hat{X}_k$.
\end{enumerate}
\end{theorem}
\proof See \ref{app:proofs}. \qed

The claim that $\hat{X}_k'=\hat{X}_k$ in Conclusion 2 of Theorem \ref{lem:newpredictionupdate} simply means that $\mbf{x}_k\in\hat{X}_k'\iff\mbf{x}_k\in\hat{X}_k$. Of course, the CZ representation of the two sets is different, which is the point. Specifically, the values of $n_{g_k}$ and $n_{c_k}$ in \eqref{eq: predUpdate set complexity - ng }--\eqref{eq: predUpdate set complexity - nc } are both reduced by $n_\text{e}$ when using $\hat{X}_k'$ relative to $\hat{X}_k$ (because both $\tilde{n}_z$ and $n_{c_p}$ are reduced by $n_\text{e}$). The additional cost of computing $\hat{X}_k'$ is discussed below. First, we complete the analysis of $\hat{X}_k'$ by showing that the additional hypothesis in Conclusion 2 of Theorem \ref{lem:newpredictionupdate} always holds when $Z$ is computed according to Assumption \ref{ass:intervalZ}, and hence we always have $\hat{X}_k'=\hat{X}_k$.

\begin{lemma} \rm \label{lem: Zr maps into Ze}
If $Z$ is computed according to Assumption \ref{ass:intervalZ}, then the reordering $(Z_\text{e},Z_\text{r}) = \mbf{U}Z$ satisfies
\begin{equation}
    \mbf{A}_\text{ee}^{-1} (\mbf{b}_\text{e} - \mbf{A}_\text{er} \mbf{z}_\text{r})\in Z_\text{e}, \quad\forall \mbf{z}_\text{r}\in Z_\text{r}.
\end{equation}
\end{lemma}
\proof See \ref{app:proofs}. \qed

\begin{corollary} \rm \label{cor: hat X' = hat X}
Let $\hat{X}_k$ and $\hat{X}_k'$ be the enclosures of $\underline{X}_k$ defined in Proposition \ref{thm:predictionupdate} and Theorem \ref{lem:newpredictionupdate}, respectively. If $Z$ is computed according to Assumption \ref{ass:intervalZ}, then $\hat{X}_k=\hat{X}_k'$.
\end{corollary}

We have established that Theorem \ref{lem:newpredictionupdate} provides an alternative CZ representation of $\hat{X}_k$ with $n_\text{e}$ fewer generators and constraints and no loss of accuracy. To compute this enclosure, the indices in $\mathcal{I}_{\text{e}}$ can be automatically stored during the construction of $P_\ell$. The computation of $\mathring{P}$, $\mbf{G}_f$, and $\mbf{c}_f$ then requires the computation of $\mbf{A}_\text{ee}^{-1}[\mbf{A}_\text{er} \ \mbf{b}_\text{e}]$, which is $n_\text{e}^2n_z$ because $\mbf{A}_\text{ee}$ is triangular, as well as the multiplications in \eqref{eq:equivalentpolytope} and \eqref{eq:gfcf}, which cost $O((n_{h_p}+n_\text{r})n_\text{e}n_z)$. Once these are known, \eqref{eq:czprnewpredictionupdate} is executed in the same way as \eqref{eq:czprpredictionupdate} (using \eqref{eq:pre_czcartprod}, \eqref{eq:czhrepintersection}, \eqref{eq:pre_czlimage}, and \eqref{eq:pre_czmsum}). The complexity of this step is the same as previously shown for \eqref{eq:czprpredictionupdate} with $n_z$ reduced by $n_\text{e}$, except that $\mbf{G}_f$ lacks the special structure of $\mbf{E}_f$. Accordingly, the linear mapping under $\mbf{G}_f$ bears the full cost of $O(n_xn_rn_g)$, where $n_g = n_{g_{k-1}}+n_{g_w}+n_{g_v}+n_r-(n_x+n_w+n_v)$ is the number of generators in $\hat{X}_{k-1} \times W \times V \times \tilde{Z}_\text{r}$.

\subsection{Proposed algorithm and implementation details}

The proposed method for set-based state estimation of nonlinear systems is given in Algorithm \ref{alg:czprestimation}.\footnote{This algorithm is implemented in ZETA \citep{Rego2025ZETA}. Available on \url{https://github.com/ZETA-Toolbox}.} The complexity reduction methods for CZs proposed in \cite{Scott2016}, denoted as $\text{red}(\cdot)$, are applied to $\hat{X}_k$ at the end of each time step $k$. 
In our implementation, Lines \ref{line: get factors of g}, \ref{line: get factors of gf}, and \ref{line: get interval Z}--\ref{line: get final polyrelax} in Algorithm \ref{alg:czprestimation} are performed automatically using a custom class that represents variables by a unique index $j$ and an interval $Z_j$. All objects in this class have access to a shared memory where $P_{\ell} = (\mbf{H}_{\ell}, \mbf{k}_{\ell}, \mbf{A}_{\ell}, \mbf{b}_{\ell})\poly$ is built. The operations described in Sections \ref{sec:arithmeticoperations}--\ref{sec:nonlinearoperations} are all implemented for objects of the class using operator overloading. When an elementary operation is evaluated with inputs of this type, a new object with a unique index $j$ is generated to represent the output and $Z_j$ is computed from the input intervals. Then, a new column $j$ is added to $\mbf{H}_{\ell}$ and $\mbf{A}_{\ell}$, and new rows are appended to $\mbf{H}_{\ell}$, $\mbf{k}_{\ell}$, $\mbf{A}_{\ell}$, and $\mbf{b}_{\ell}$ encoding $Q_j$. These objects can be passed through arbitrary factorable functions implemented in standard code to obtain $Z$ and $P_\ell$. Importantly, the user does not need to specify the factorable representation. This makes obtaining the polyhedral relaxation straightforward, without requiring specific implementations for each function investigated.

\begin{algorithm}[!htb] 
	\caption{CZPR for state estimation of \eqref{eq:systemequations}.}
	\label{alg:czprestimation}
	\small
	\begin{algorithmic}[1]
        \State Let $(X_0,W,V) \subset \realset^{n_x} \times \realset^{n_w} \times \realset^{n_v}$ be constrained zonotopes, and $(\mbf{u}_k,\mbf{y}_k) \in \realset^{n_u} \times \realset^{n_y}$, $k \in [0,N]$. These variables must satisfy $(\mbf{x}_0,\mbf{w}_k,\mbf{v}_k) \in X_0 \times W \times V$ and \eqref{eq:systemg};
        \State Obtain a factorable representation of the function $\mbf{g}(\mbf{x}_k,\mbf{v}_{k})$; \label{line: get factors of g}
        \State Compute $\hat{X}_0$ according to Remark \ref{rem:czprinitialupdate};
        \State Obtain a factorable representation of the composite function $\bm{\ell} (\mbf{x}_{k-1},\mbf{w}_{k-1},\mbf{v}_k,\mbf{u}_{k-1}) = \mbf{g}(\mbf{f}(\mbf{x}_{k-1},\mbf{w}_{k-1},\mbf{u}_{k-1}),\mbf{v}_k)$; \label{line: get factors of gf}
  	    \For {$k = 1,\ldots,N$}
        \State $S \gets \hat{X}_{k-1} \times W \times V$;
        \State Compute $Z$ as in Remark \ref{rem:intervalZ} with $\square S$ as input; \label{line: get interval Z}%
        \State Compute $Q_j$ for each $\alpha_j$, $j>n_x+n_w+n_v$, using $Z$;  \label{line: compute halfspaces}
        \State $P_\ell \gets \bigcap_{j=n_x+n_w+n_v+1}^{n_z} Q_j$ using \eqref{eq:hrepintersection}; \label{line: get final polyrelax}
        \State $P_y \gets (\noarg, \noarg, \mbf{E}_\ell, \mbf{y}_k)\poly$;
        \State Compute $P_\ell \cap P_y$ using \eqref{eq:hrepintersection};
        \State Obtain $\mathring{P} = (\mathring{\mbf{H}},\mathring{\mbf{k}},\mathring{\mbf{A}},\mathring{\mbf{b}})\poly$, $\mbf{G}_f$, and $\mbf{c}_f$ using \eqref{eq:equivalentpolytope} and \eqref{eq:gfcf};
        \State Obtain $\tilde{Z}_\text{r}$ from $Z$;        
        \State $\hat{X}_k \gets \mbf{G}_f ((\hat{X}_{k-1} \times W \times V \times \tilde{Z}_\text{r}) \cap \mathring{P}) \oplus \mbf{c}_f$;
        \State $\hat{X}_k \gets \text{red}(\hat{X}_k)$;
        \EndFor
	\end{algorithmic}
	\normalsize
\end{algorithm}	

\section{Numerical examples} \label{sec:results}

This section illustrates the advantages of Algorithm \ref{alg:czprestimation} (CZPR) compared to: (i) the CZ-based mean value extension in \cite{Rego2021} (CZMV), with heuristic \emph{C2} therein; and (ii) the CZ enclosures obtained by DC programming in \cite{DePaula2024DC} (CZDC), using Prop.~4 therein for DC function decomposition, and using the interval hulls of the CZ enclosures for vertex evaluations. Numerical simulations were performed using MATLAB 9.1 with Gurobi 10.0.1 for solving LPs, while computational times are measured on a laptop with 32 GB RAM and an Intel Core Ultra 9 185H processor. For all methods, the number of constraints and generators in the CZs is limited using the reduction methods in \cite{Scott2016}. Before these simulations, we present a short example demonstrating the advantages and disadvantages of CZPR in comparison to hybrid zonotopes \citep{Siefert2025HZFuncDecomp}, denoted by HZ.

Computing the exact volumes of CZs is intractable. Thus, enclosure size is compared using the $n_x^\text{th}$ roots of $\intvol{\hat{X}_k}$ and $\parvol{\hat{X}_k}$, which are the exact volumes of the interval hull $\square \hat{X}_k$ and a parallelotope enclosure $\diamond \hat{X}_k$. The latter is obtained using Prop.~3 in \cite{DePaula2024DC}, initialized with the parallelotope obtained by reducing $\hat{X}_k$ to an order-one zonotope through constraint elimination and generator reduction. We also define the geometric average volume ratio $\intavr$ (resp.~$\paravr$) as the ratio of $\sqrt[n_x]{\intvol{\hat{X}_k}}$ (resp.~$\sqrt[n_x]{\parvol{\hat{X}_k}}$) for one method over that of another, geometrically averaged over all time steps.

\subsection{Example 1} \label{ex:HZcomparison}

Consider a nonlinear system with dynamics $x_{1,k} = x_{1,k-1} + T_s x_{2,k-1}$ and $x_{2,k} = x_{2,k-1} + T_s (10 \text{sin}(x_{1,k-1}))$, with $T_s = 0.1$s, adapted from zonoLAB \citep{Koeln2024zonoLAB}. The initial state set $X_0$ is %
\begin{equation} \label{eq:exampleHZX0}
X_0 = \left( \begin{bmatrix} \frac{\pi}{6} & 0 \\ 0 & 0.1 \end{bmatrix}, \begin{bmatrix} \frac{\pi}{3} \\ 0 \end{bmatrix} \right)\zon.
\end{equation}

\begin{figure}[!thb]
	\centering{
		\def\svgwidth{0.8\columnwidth}
        {\small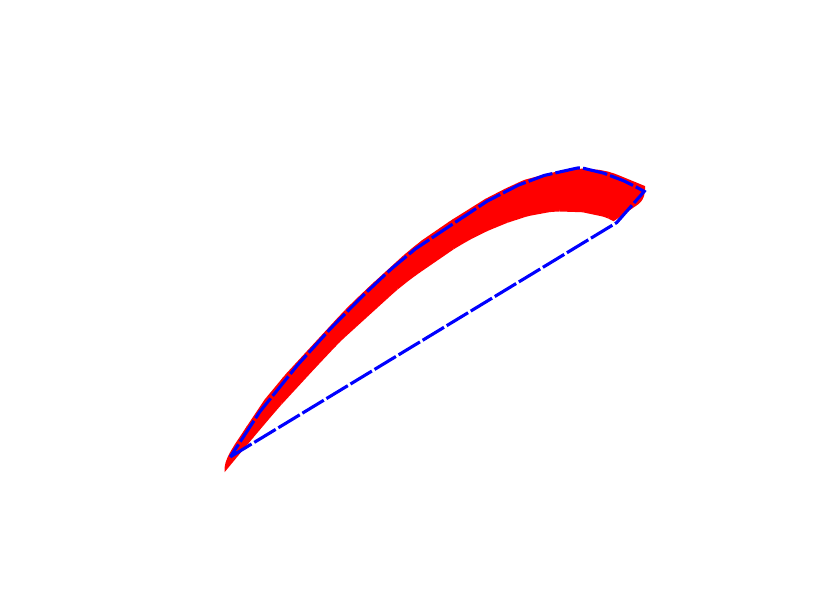}
		\caption{The state enclosures computed for $k=5$ using HZ (solid red) and CZPR (dashed blue) in Example 1.}\label{fig:examplehybzono}}
\end{figure}

In this example, we compute the prediction set $\hat{X}_k^p$ of CZPR, separately as described in Remark \ref{rem:separatesteps}, with 
$\hat{X}_0^p \triangleq X_0$, without order reduction methods. The set computed with hybrid zonotopes is obtained using the implementation available in the \code{exampleNonlinear.m} file in zonoLAB, adapted to this example while maintaining all the design parameters. Figure \ref{fig:examplehybzono} shows the propagated sets for $k = 5$ using HZ (solid red) and CZPR (dashed blue). Thanks to their ability of describing non-convex sets, HZ generates a less conservative enclosure than CZPR. However, as it is shown in Table \ref{tab:hybzonocomplexity} ($n_g$, $n_c$, and $n_b$ stand for the numbers of generators, constraints and binary variables, respectively), this feature comes with a heavy drawback. Although also with linear complexity growth over time, HZ presents substantially higher set complexity than CZPR. Combined with the absence of efficient order reduction methods in the current literature, HZ can become intractable with a few dozens of iterations in this simple example. Due to this reason, in the remainder of this section, we focus on comparing CZPR with CZMV and CZDC in more complex examples.

\begin{table}[!htb]
    \footnotesize
	\centering
	\caption{Complexity of the propagated sets over time for Example 1.}
	\begin{tabular}{c | c | c  c c c c c } \hline
        \multicolumn{2}{r|}{$k$} & 0 & 1 & 2 & 3 & 4 & 5 \\ \hline
        \multirow{2}{*}{CZPR} & $n_g$ & 2 & 14 & 27 & 40 & 53 & 66 \\
                              & $n_c$ & 0 & 11 & 23 & 35 & 47 & 59 \\ \hline
        \multirow{3}{*}{HZ}   & $n_g$ & 2 & 93 & 184 & 275 & 366 & 457 \\
                              & $n_c$ & 0 & 49 & 98  & 147 & 196 & 245 \\
                              & $n_b$ & 0 & 42 & 84  & 126 & 168 & 210 \\ \hline
	\end{tabular}    
    \normalsize
	\label{tab:hybzonocomplexity}   
\end{table}

\subsection{Example 2} \label{ex:simple}

Consider the following system from \cite{Rego2021}:
\begin{align*}%
x_{1,k} & = 3 x_{1,k-1} - \frac{x_{1,k-1}^2}{7} - \frac{4 x_{1,k-1} x_{2,k-1}}{4 + x_{1,k-1}} + w_{1,k-1}, \\ x_{2,k} & = -2 x_{2,k-1} + \frac{3 x_{1,k-1} x_{2,k-1}}{4 + x_{1,k-1}} + w_{2,k-1}, \\
y_{1,k} & =  x_{1,k} - \sin\left(\frac{x_{2,k}}{2}\right) + v_{1,k}, \quad
y_{2,k} = -x_{1,k}x_{2,k} + x_{2,k} + v_{2,k},
\end{align*}
with $\|\mbf{w}_k\|_\infty \leq 0.8$, $\|\mbf{v}_k\|_\infty \leq 0.4$, and
\begin{equation} \label{eq:example1X0}
X_0 = \left( \begin{bmatrix} 0.5 & 1 & -0.5 \\ 0.5 & 0.5 & 0 \end{bmatrix}, \begin{bmatrix} 5 \\ 0.5 \end{bmatrix} \right)\zon.
\end{equation}
The initial state is $\mbf{x}_0 = (5.2,0.65)$ for simulations.

\begin{figure}[!thb]
	\centering{
		\def\svgwidth{\columnwidth}
  {\scriptsize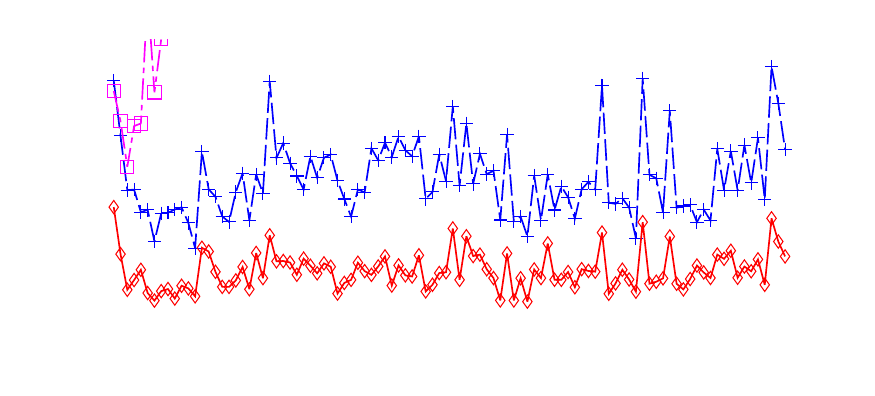}
		\def\svgwidth{\columnwidth}
  {\scriptsize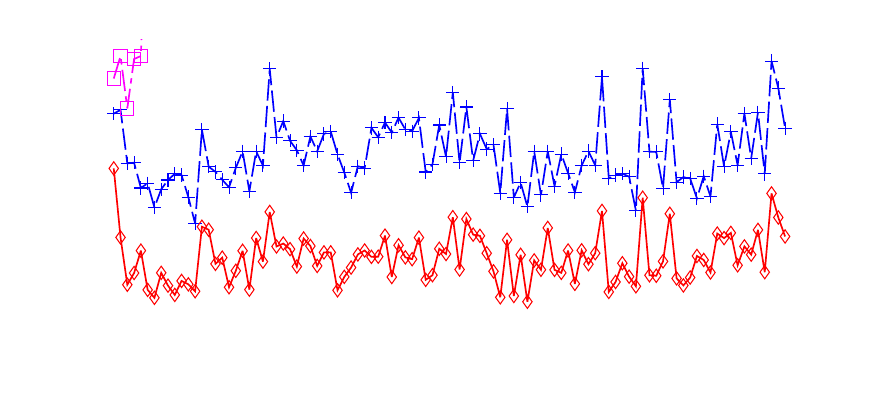}
		\caption{$\sqrt[n_x]{\intvol{\hat{X}_k}}$ (top) and $\sqrt[n_x]{\parvol{\hat{X}_k}}$ (bottom) for CZMV, CZDC, and CZPR for Example 2.}\label{fig:example1volume}}%
\end{figure}

\begin{figure}[!thb]
	\centering{
		\def\svgwidth{0.7\columnwidth}
        {\small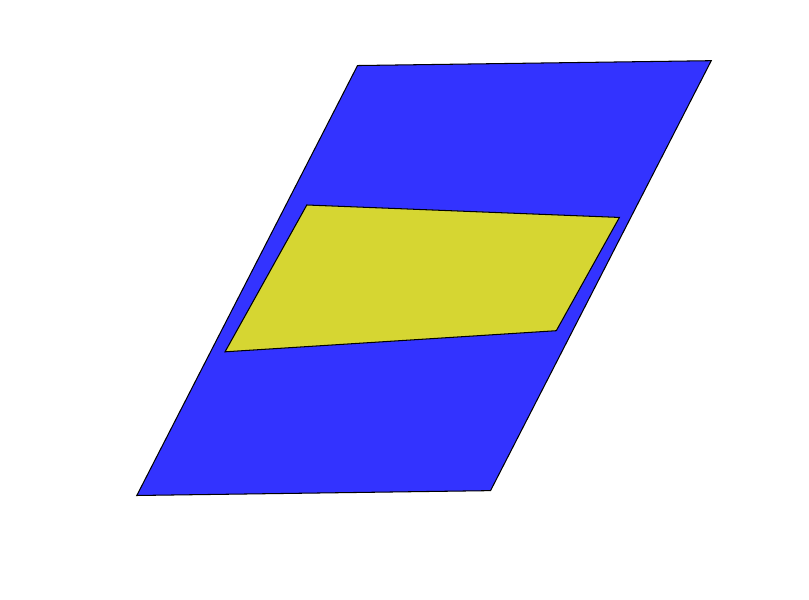}
        \caption{The enclosure $\hat{X}_{40}$ from CZMV (blue) and CZPR (yellow) for Example 2.}\label{fig:example1w08projection}}
\end{figure}

Figure \ref{fig:example1volume} shows $\sqrt[n]{\intvol{\hat{X}_k}}$ and $\sqrt[n]{\parvol{\hat{X}_k}}$ for the enclosures $\hat{X}_k$ obtained using CZMV, CZDC, and CZPR, for $k \in [0, 100]$. The numbers of generators and constraints were limited to 20 and 8, respectively. The enclosures provided by CZDC diverge quickly, while those produced by both CZMV and CZPR remain bounded. However, CZPR provides notably tighter enclosures than CZMV, with $49.79\%$ and $53.38\%$ CZPR-to-CZMV $\intavr$ and $\paravr$, respectively. The average computational times per $k$ to compute $\hat{X}_k$ prior to order reduction were 21.9 ms for CZMV and 15.3 ms for CZPR, while the average order reduction times per $k$ were 10.9 ms and 56.2 ms, respectively. CZPR was faster than CZMV in the computation of the state enclosures in this example, but with slower total times when considering the order reduction. This highlights that the polyhedral relaxation of $\bm{\ell}$ leads to a more complex enclosure than the one obtained by CZMV, requiring more computational effort by the order reduction method. Without the complexity reduction step proposed in Sec.~\ref{sec:equivalentset}, the average order reduction time per $k$ of CZPR was 87.5 ms. This demonstrates the importance of using \eqref{eq:czprnewpredictionupdate} for reduced computational cost of CZPR. Figure \ref{fig:example1w08projection} compares the sets $\hat{X}_{40}$ provided by CZMV and CZPR, illustrating the difference in conservatism for this example.

\subsection{Example 3} \label{ex:chemicalreactor}
Consider the following chemical reactor model with linear measurements adapted and discretized from Example 4.5.4 in \cite{ScottPhdThesis}:
\begin{equation*}
    \mbf{x}_k = \mbf{x}_{k-1} + T_s \mbf{S} \mbf{r}(\mbf{x}_{k-1},\mbf{w}_{k-1}), \quad \mbf{y}_k = \mbf{C} \mbf{x}_k + \mbf{v}_k,
\end{equation*}
where
\begin{equation*}
    \mbf{S} {\triangleq} \begin{bmatrix}
        -1 &  0 &  0 &  0 &  0 \\
        -1 &  0 &  0 &  0 &  0 \\
         1 & -1 & -1 &  0 & -2 \\
         1 &  0 &  0 &  0 &  0 \\
         0 & -1 & -1 &  0 &  0 \\
         0 &  1 &  0 &  0 &  0 \\
         0 &  0 &  1 & -1 &  0 \\
         0 &  0 &  0 &  1 &  0 \\
         0 &  0 &  0 &  1 &  0 \\
         0 &  0 &  0 &  0 &  1 
    \end{bmatrix},  \begin{aligned} \mbf{r}(\mbf{x},\mbf{w}) & \triangleq \begin{bmatrix}
    w_1 x_1 x_2 \\
    w_2 (x_3 x_5 - (1/\kappa_1) x_6) \\
    w_3 (x_3 x_5 - (1/\kappa_2) x_7) \\
    w_4 x_7 \\
    w_5 x_3^2
    \end{bmatrix}, \\
    \mbf{C} & \triangleq \begin{bmatrix}
    \eye{8} & \ones{8}{2}
    \end{bmatrix}.
    \end{aligned}
\end{equation*}
The parameters are $\kappa_1 = 2081$, $\kappa_2 = 4162$, and $T_s = 0.01$. The uncertainties are bounded by $w_{1,k} \in [23,83]$, $w_{2,k} \in [100,600]$, $w_{3,k} \in [100,600]$, $w_{4,k} \in [0.001,50]$, $w_{5,k} \in [900,1500]$, $\|\mbf{v}_k\| \leq 0.006$, and $X_0 = ( 10^{-5} {\cdot} \eye{10}, \mbf{x}_0)\zon$ with $\mbf{x}_0 = (1.53\ten{-4},0.4,0,0,1.9\ten{-3},\zeros{5}{1})$. The initial condition $\mbf{x}_0$ was used for all simulations.

\begin{figure}[!thb]
	\centering{
		\def\svgwidth{\columnwidth}
  {\scriptsize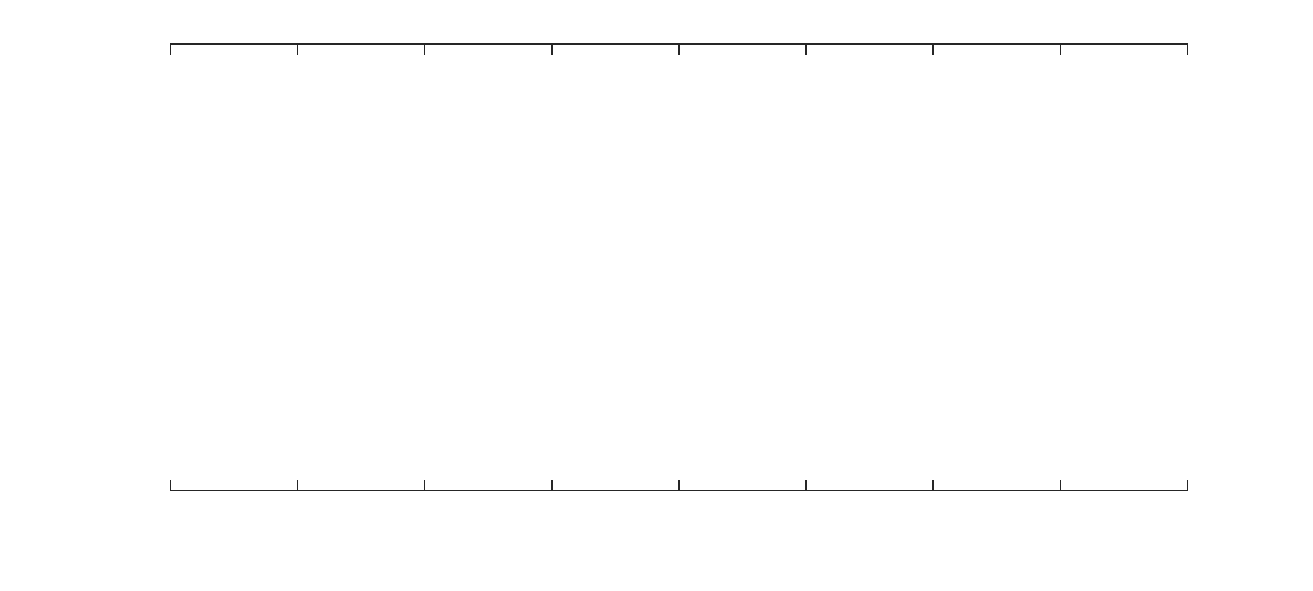}
		\def\svgwidth{\columnwidth}
  {\scriptsize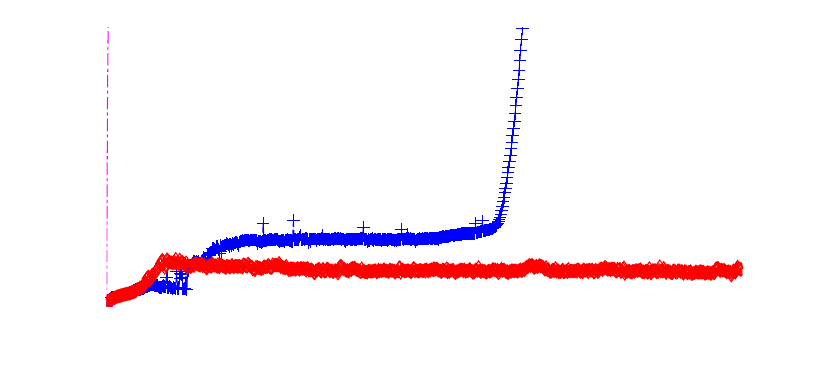}
		\caption{$\sqrt[n_x]{\intvol{\hat{X}_k}}$ (top) and $\sqrt[n_x]{\parvol{\hat{X}_k}}$ (bottom) for CZMV, CZDC, and CZPR for Example 3.}\label{fig:example2volume}}
\end{figure}

Figure \ref{fig:example2volume} shows $\sqrt[n_x]{\intvol{\hat{X}_k}}$ and $\sqrt[n_x]{\parvol{\hat{X}_k}}$ for CZMV, CZDC and CZPR, with $k \in [0, 800]$. The numbers of generators and constraints were limited to 80 and 30, respectively. CZDC diverges in the very first iterations. Additionally, while CZMV provides more accurate enclosures at the beginning, it gets more conservative after some iterations until it diverges as well. On the other side, CZPR remains bounded. The average computational times per $k$ to compute $\hat{X}_k$ prior to order reduction were 163.2 ms for CZMV and 31.7 ms for CZPR, while the average order reduction times per $k$ were 281 ms and 249.3 ms, respectively. CZPR was considerably faster for this example because the factorable representation of the system dynamics is relatively simple compared to Example 2. Without the reduction step in Sec.~\ref{sec:equivalentset}, the average computational time was 589.3 ms. 

\subsection{Example 4} 

\begin{figure}[htb]
	\centering{
		\includegraphics[width=0.6\columnwidth]{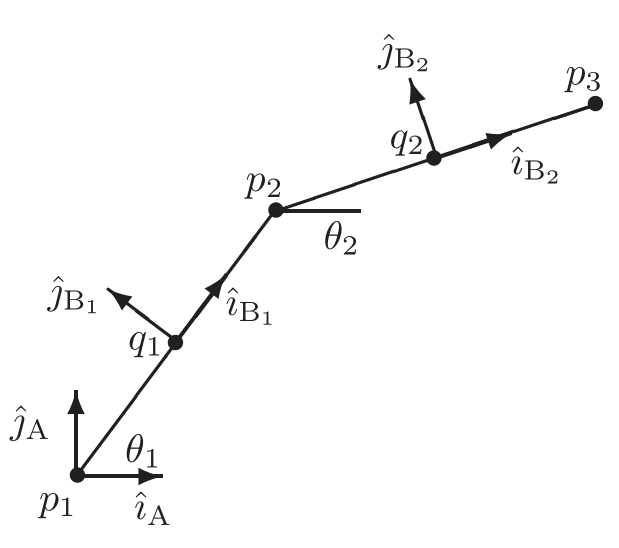}
		\caption{Planar robotic arm.}
		\label{fig:example_planararm}}
\end{figure}

Consider a planar robotic arm with revolute joints $p_1$ and $p_2$ and end effector $p_3$ as shown in Figure \ref{fig:example_planararm}. The points $q_1$ and $q_2$ are the centers of mass of the two links. After Euler discretization, the dynamics are \citep{Damato2011}:
\begin{equation*}
	\mbf{x}_k = \mbf{x}_{k-1} + T_s\begin{bmatrix} x_{3,k-1} \\ x_{4,k-1} \\ \mbf{M} (\mbf{x}_{k-1})^{-1} ( - \bm{\gamma}(\mbf{x}_{k-1}) + \mbf{q}(u_{k-1}) )\end{bmatrix},
\end{equation*}
where $\mbf{q}(u) \triangleq (u,0)$, $\bm{\gamma}(\mbf{x}) \triangleq (\gamma_1(\mbf{x}), \gamma_2(\mbf{x}))$, $\gamma_1(\mbf{x}) \triangleq \frac{1}{2} m_2 l_1 l_2 \sin(x_1 - x_2) x_4^2 + (k_1 + k_2) x_1 - k_2 x_2 + (c_1 + c_2) x_3 - c_2 x_4$, $\gamma_2(\mbf{x}) \triangleq -\frac{1}{2} m_2 l_1 l_2 \sin(x_1 - x_2) x_3^2 - k_2 x_1 + k_2 x_2 - c_2 x_3 + c_2 x_4$, 
\begin{equation*}
  \mbf{M}(\mbf{x}) \triangleq \begin{bmatrix} \frac{1}{3}m_1 l_1^2 + m_2 l_1^2 & \frac{1}{2} m_2 l_1 l_2 \cos(x_{1} - x_{2}) \\ \frac{1}{2} m_2 l_1 l_2 \cos(x_{1} - x_{2}) & \frac{1}{3} m_2 l_2^2 \end{bmatrix}.
\end{equation*}
Above, $l_1 = 3$ m, $l_2 = 2$ m, $m_1 = 2$ kg, and $m_2 = 1$ kg are the lengths and masses of the links; $c_1 = 10$ N${\cdot}$m/(rad/s) and $c_2 = 1$ N${\cdot}$m/(rad/s) are friction coefficients; and $k_1 = 7$ N${\cdot}$m/rad and $k_2 = 5$ N${\cdot}$m/rad are stiffness coefficients. The input $u_k = 20 \sin(kT_s)$ N${\cdot}$m is the torque applied to the first joint. Moreover, $x_1 \triangleq \theta_1$ and $x_2 \triangleq \theta_2$ are the angles of the links with respect to the inertial frame, and $x_3 \triangleq \dot{\theta}_1$ and $x_4 \triangleq \dot{\theta}_2$. We consider a challenging scenario in which only the position of the end effector is measured, i.e.,
\begin{equation*}
    \mbf{y}_k = \begin{bmatrix} \cos(x_{1,k}) l_1 + \cos(x_{2,k}) l_2 \\ \sin(x_{1,k}) l_1 + \sin(x_{2,k}) l_2 \end{bmatrix} + \mbf{v}_k,
\end{equation*}
with $\|\mbf{v}_k\|_\infty \leq 0.01$. The initial state is $\mbf{x}_0 = \zeros{4}{1}$ and
\begin{equation*}
    X_0 = \left(\begin{bmatrix} 0.1745 & 0 & 0 & 0 \\ 0 & 0.1745 & 0 & 0  \\ 0 & 0 & 0.0873 & 0 \\ 0 & 0 & 0 & 0.0873 \end{bmatrix}, \zeros{4}{1} \right)\zon.
\end{equation*}

\begin{figure}[!thb]
	\centering{
		\def\svgwidth{\columnwidth}
  {\scriptsize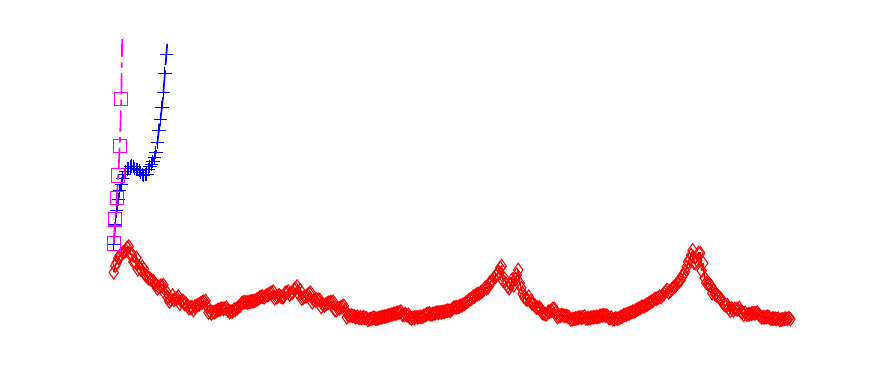}
		\def\svgwidth{\columnwidth}
  {\scriptsize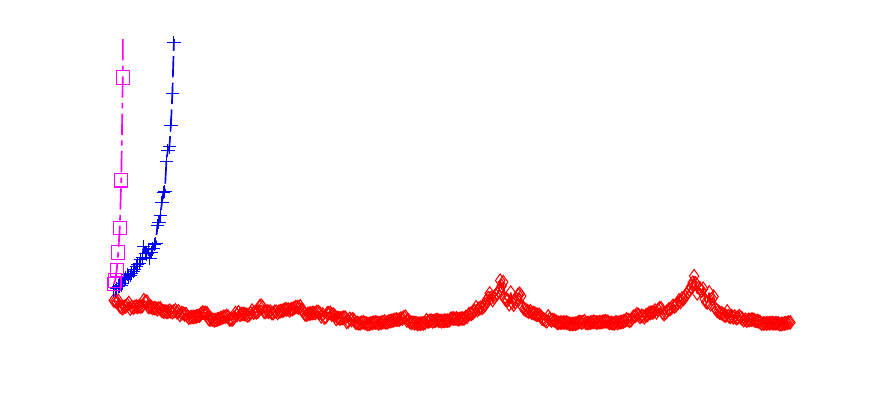}
		\caption{$\sqrt[n_x]{\intvol{\hat{X}_k}}$ (top) and $\sqrt[n_x]{\parvol{\hat{X}_k}}$ (bottom) for CZMV, CZDC, and CZPR for Example 4.}\label{fig:example3volume}}
\end{figure}

Figure \ref{fig:example3volume} shows $\sqrt[n_x]{\intvol{\hat{X}_k}}$ and $\sqrt[n_x]{\parvol{\hat{X}_k}}$ for the enclosures $\hat{X}_k$ obtained using CZMV, CZDC, and CZPR with $k \in [0, 450]$. The numbers of generators and constraints were limited to 60 and 20, respectively. Both CZDC and CZMV diverge quickly, while only CZPR remains bounded. The average computational times per $k$ for CZPR to compute $\hat{X}_k$ prior to order reduction were 31.8 ms, while the average order reduction times per $k$ were 1.56 s. This is a significant cost and was dominated by extensive reduction due to high-complexity CZs ($\hat{X}_k$ had $\sim$205 generators and $\sim$146 constraints before reduction). In addition to the factorable representation having many more factors than in the previous examples, the $Q_j$ for the trigonometric functions here contribute a substantial number of halfspaces to $P_{\ell}$. Since these functional forms are common in robotic systems, further efficiency improvements should be investigated in future work. Still, CZPR was the only method able to provide effective enclosures.

Moreover, without the lower-complexity enclosure developed in Sec~\ref{sec:equivalentset}, the average order reduction time increases significantly to 3.97 s ($\hat{X}_k$ had $\sim$275 generators and $\sim$215 constraints before reduction). Interestingly, the additional reduction also led to significant additional conservatism for this example, ultimately causing the enclosures to diverge. We conclude that the standard reduction methods cannot be relied upon to find the exact reductions identified in Sec.~\ref{sec:equivalentset}, which further highlights the utility of the algorithmic improvements there.

\section{Conclusions} \label{sec:conclusions}

This paper developed a novel method, CZPR, for set-based state estimation of nonlinear discrete-time systems using polyhedral relaxations and constrained zonotopes. CZPR avoids the need to linearize nonlinear mappings and bound linearization errors over potentially large input set. Thanks to this fact, it is able to generate state enclosures that are less conservative than existing state-of-the-art CZ methods based on the Mean Value Theorem and DC programming, as highlighted in three numerical examples. Future work will address the reduction of computational effort regarding functional forms for which CZPR is able to provide good enclosures, but with considerably high execution times.

\appendix

\section{Polyhedral enclosure of the sine function} \label{app:sine}

\begin{figure}[!thb]
	\centering{
		\def\svgwidth{\columnwidth}
        {\small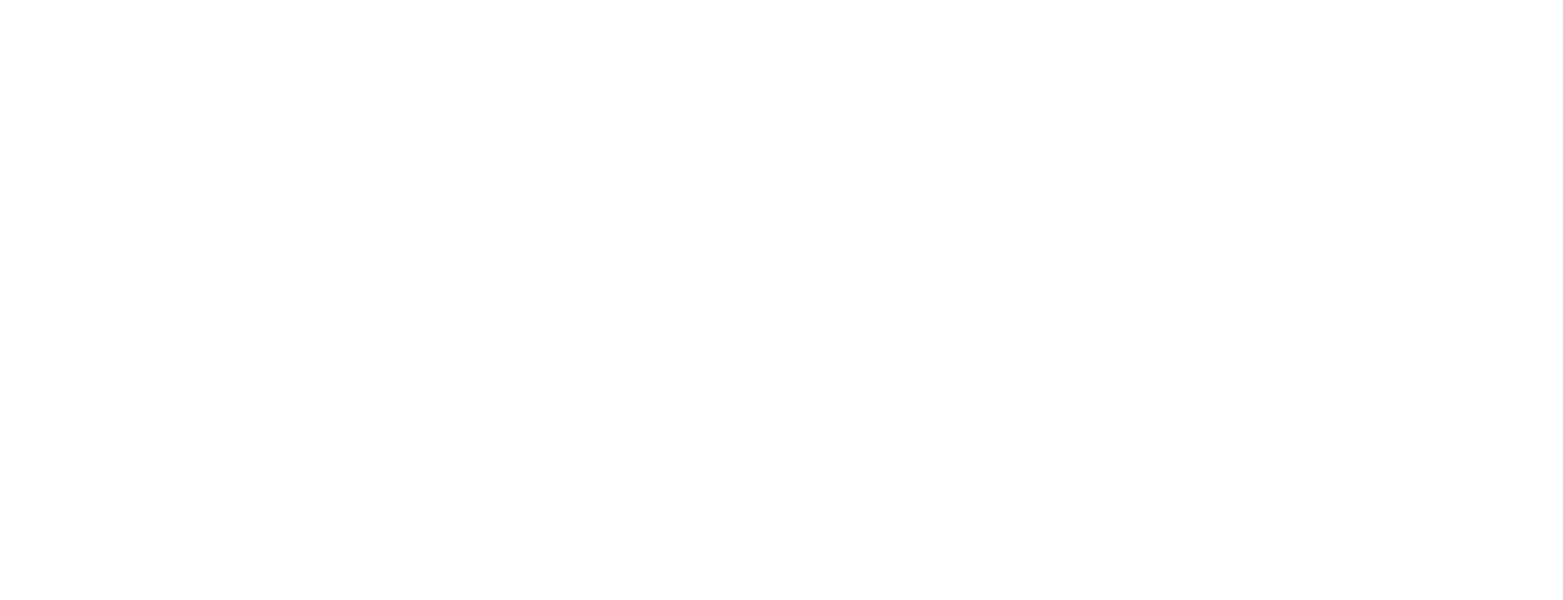}
		\caption{Convex envelope for $\text{sin}(z_a)$ on the intervals $[\frac{3\pi}{2},~\frac{6\pi}{2}]$ (left) and $[\frac{9\pi}{4},~\frac{7\pi}{2}]$ (right).}\label{fig:sin_twoexamples}}
\end{figure}

We first present the convex underestimator $\betacv(z_a)$ for $z_j = \beta_j(z_a) \triangleq \text{sin}(z_a)$, $z_a \in Z_a \triangleq [\lb{z}_a, \ub{z}_a]$, based on the ideas in \cite{ScottPhdThesis}. The underestimator is developed for $Z_a \subset [\frac{3\pi}{2},~\frac{7\pi}{2}]$, then extended to an arbitrary input interval. Even for $Z_a \subset [\frac{3\pi}{2},~\frac{7\pi}{2}]$, there are several cases to consider. To illustrate the possibilities, Figure~\ref{fig:sin_twoexamples} shows the convex envelopes for two different input intervals contained in $[\frac{3\pi}{2},~\frac{7\pi}{2}]$.

In the first case (left), the dashed vertical line represents a special point denoted $z_a^{\text{S}_1}$. The envelope consists of a nonlinear portion to the left of $z_a^{\text{S}_1}$, where it equals $\sin(z_a)$, and a linear portion to the right of $z_a^{\text{S}_1}$, where it is the secant between $z_a^{\text{S}_1}$ and $\ub{z}_a$. The point $z_a^{\text{S}_1}$ is located such that the secant has the same slope as $\sin(z_a)$ at $z_a^{\text{S}_1}$. 
Notably, the lower nonlinear portion does not always exist. For example, the envelope would become a pure secant if $\lb{z}_a$ were moved to the right of $z_a^{\text{S}_1}$.

The second case in Figure~\ref{fig:sin_twoexamples} (right) is analogous, but shows the possible existence of an upper nonlinear portion as well, to the right of some $z_a^{\text{S}_2}$. In general, we must consider the possibility of three \emph{subregions}, denoted by $[\lb{z}_a,\lb{\rho}]$ (lower nonlinear region), $[\lb{\rho},\ub{\rho}]$ (secant region), and $[\ub{\rho},\ub{z}_a]$ (upper nonlinear region). We take $\lb{\rho}=\lb{z}_a$ (resp.~$\ub{\rho}=\ub{z}_a$) if the lower (resp.~upper) subregion does not exist, as in the right (resp.~left) panel of Figure~\ref{fig:sin_twoexamples}. Note that the only case where all three subregions exist is when $Z_a=[\frac{3\pi}{2},\frac{7\pi}{2}]$, in which the lower and upper regions are singletons at $\frac{3\pi}{2}$ and $\frac{7\pi}{2}$, respectively. With these definitions, the convex envelope on any $Z_a \subset [\frac{3\pi}{2},~\frac{7\pi}{2}]$ is
\begin{equation*} %
\small
    \eta(Z_a,z_a) \triangleq \left\{\begin{aligned}
        \sin(z_a) & ~ \text{ if } ~ z_a \in [\lb{z}_a, \lb{\rho}],\\ 
        \sin(\lb{\rho}) + \frac{\sin(\ub{\rho}) - \sin(\lb{\rho})}{\ub{\rho}-\lb{\rho}}(z_a - \lb{\rho}) & ~ \text{ if } ~ z_a \in [\lb{\rho}, \ub{\rho}],\\ 
        \sin(z_a) & ~ \text{ if } ~ z_a \in [\ub{\rho}, \ub{z}_a]. 
        \end{aligned} \right.
\end{equation*}  

The general procedure for finding $\lb{\rho}$ and $\ub{\rho}$ is as follows:

\begin{itemize}
\item If $\ub{z}_a \leq 2\pi$, then $\sin(z_a)$ is convex on $Z_a$ and only the lower subregion exists. Set $\lb{\rho}=\ub{\rho}=\ub{z}_a$.
\item If $\lb{z}_a \geq 3\pi$, then $\sin(z_a)$ is convex on $Z_a$ and only the upper subregion exists. Set $\lb{\rho}=\ub{\rho}=\lb{z}_a$.
\item If $\ub{z}_a > 2\pi$ and $\lb{z}_a < 3\pi$:
\begin{itemize}
\item Let $z^{\text{S}_1}_a$ be the unique solution of $\sin(\ub{z}_a) - \sin(z_a) = (\ub{z}_a - z_a) \cos(z_a)$ with $z_a \in [\frac{3\pi}{2}, 2\pi]$. This is the point where the slope of the secant from $\ub{z}_a$ matches the slope of $\sin{z_a}$. It cannot lie above the inflection point at $2\pi$ and can be found by bisection on $[\frac{3\pi}{2}, 2\pi]$.
\item The lower subregion exists if $z^{\text{S}_1}_a>\lb{z}_a$ and not otherwise. In either case, set $\lb{\rho} = \max(\lb{z}_a,z^{\text{S}_1}_a)$.
\item Let $z^{\text{S}_2}_a$ be the solution of $\sin(z_a) - \sin(\lb{z}_a) = (z_a - \lb{z}_a)\cos(z_a)$ for $z_a \in [3\pi, \frac{7\pi}{2}]$. This is the point where the slope of the secant from $\lb{z}_a$ matches the slope of $\sin{z_a}$. It cannot lie below the inflection point at $3\pi$ and can be found by bisection on $[3\pi, \frac{7\pi}{2}]$.
\item The upper subregion exists if $z^{\text{S}_2}_a<\ub{z}_a$ and not otherwise. In either case, set $\ub{\rho} = \min(\ub{z}_a,z^{\text{S}_2}_a)$.
\end{itemize}
\end{itemize}

We can now develop a polyhedron $\mathcal{Q}^\text{CV}_\eta(Z_a)$ that encloses the epigraph of $\eta(Z_a,\cdot)$ on any $Z_a \subset [\frac{3\pi}{2},~\frac{7\pi}{2}]$ as follows:
\begin{itemize}
    \item If $\lb{z}_a<\lb{\rho}$, then add three constraints to $\mathcal{Q}^\text{CV}_\eta(Z_a)$ requiring $z_j$ to lie above the linearizations of $\eta(Z_a,\cdot)$ at $\lb{z}_a$, $\text{mid}([\lb{z}_a, \lb{\rho}])$, and $\lb{\rho}$. Since $\eta(Z_a,\cdot)=\sin$ at these points, the constraint for linearization point $y$ is $z_j \geq \cos(y) (z_a - y) + \sin(y)$.
    \item If $\lb{\rho} < \ub{\rho}$, then add the constraint $z_j \geq \sin(\lb{\rho}) + \frac{\sin(\ub{\rho}) - \sin(\lb{\rho})}{\ub{\rho}-\lb{\rho}}(z_a - \lb{\rho})$.
    \item If $\ub{\rho}<\ub{z}_a$, then add three constraints requiring $z_j$ to lie above the linearizations of $\eta(Z_a,\cdot)$ at $\ub{\rho}$, $\text{mid}([\ub{\rho},\ub{z}_a])$, and $\ub{z}_a$.
\end{itemize}

\begin{figure}[!thb]
	\centering{
		\def\svgwidth{0.6\columnwidth}
        {\small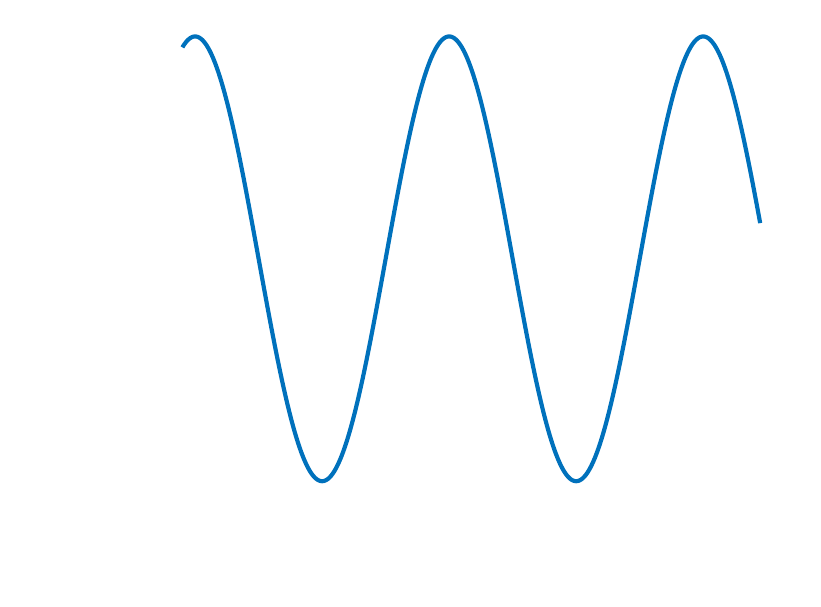}
		\caption{Convex envelope for $\text{sin}(z_a)$ on the interval $[-3.6\pi,~0.95\pi]$.}\label{fig:sin_multi}}
\end{figure}

To construct a convex understimator for a general $Z_a$, we must consider the possibility that many periods are involved. The convex envelope for such a case is shown in Figure~\ref{fig:sin_multi}. The envelope consists of a constant portion in the middle, where it is $-1$, and two end portions that are each contained within a single `trough-to-trough' period of the form $\Gamma(n)\triangleq[\frac{3\pi}{2}+2\pi n,~\frac{7\pi}{2}+2\pi n]$. Thus, the complete relaxation is described by considering these three \emph{regions} of $Z_a$ (not to be confused with the \emph{subregions} used to define $\eta$ above).

Note that each $\Gamma(n)$ is an $n$-period translation of the base interval $[\frac{3\pi}{2},~\frac{7\pi}{2}]$ considered above. Thus, the envelopes on these regions can be obtained by applying the function $\eta$ defined above with suitably translated arguments. To find the required translations, define $\mathcal{N}(z_a) \triangleq \frac{1}{2\pi} z_a - \frac{3}{4}$ and note that this function maps $\Gamma(n)$ onto $[n,n+1]$. Therefore, for any $z_a$, the value $n=\text{floor}(\mathcal{N}(z_a))$ provides the index of a $\Gamma(n)$ to which $z_a$ belongs, and we can infer that $z_a$ is bracketed between the $n^\text{th}$ and $(n+1)^\text{st}$ troughs,
\begin{equation}
\frac{3\pi}{2}+2\pi n\leq z_a < \frac{3\pi}{2}+2\pi(n+1).
\end{equation}
The strict inequality occurs because floor will associate the point on the boundary between $\Gamma(n)$ and $\Gamma(n+1)$ with the index $n$, not $n+1$. The alternative value $n'=\text{ceil}(\mathcal{N}(z_a))-1$ (which will only differ from $n'$ if $z_a$ is a boundary point) satisfies the same bounds, but with the first inequality strict and the second weak.

Letting $\lb{n}\triangleq \text{ceil}(\mathcal{N}(\lb{z}_a))-1$ and $\ub{n}\triangleq \text{floor}(\mathcal{N}(\ub{z}_a))$, we may now partition $Z_a$ into the three regions discussed above:
\begin{itemize}
\item Region 1: $Z_a^1\triangleq Z_a\cap\Gamma(\lb{n})$. This region runs from $\lb{z}_a$ to the next trough or to $\ub{z}_a$, whichever comes first.
\item Region 2: $Z_a^2\triangleq \cup_{\lb{n}<n<\ub{n}}\Gamma(n)$. This region covers all periods $\Gamma(n)$ that lie entirely in the interior of $Z_a$. 
\item Region 3: $Z_a^3\triangleq Z_a\cap\Gamma(\ub{n})$. This region ends at $\ub{z}_a$ and extends backwards to the nearest trough or to $\lb{z}_a$, whichever comes first.
\end{itemize}
The differing use of floor and ceil for $\lb{n}$ and $\ub{n}$ ensures that $Z_a^1$ and $Z_a^3$ are never full periods (e.g., $Z_a^1\neq\Gamma(\lb{n})$). Instead, if $\lb{z}_a$ (resp.~$\ub{z}_a$) is a trough, then $Z_a^1$ (resp.~$Z_a^3$) will be a singleton that is already contained in $Z_a^2$, and so can be taken to be empty. Similarly, if $\lb{n}=\ub{n}$, then $Z_a^1=Z_a^3$ and we can take $Z_a^3$ to be empty for the purpose of defining a piecewise relaxation. Finally, note that $Z_a^2$ can be empty.

Define the $n$-period translation $t_n(z_a)\triangleq z_a-2\pi n$, so that $t_n(\Gamma(n))=\Gamma(0)=[\frac{3\pi}{2},\frac{7\pi}{2}]$. Then, the convex envelope on $Z_a$ is
\begin{equation}  \label{eq:sinconvexest}
    \betacv(z_a) \triangleq \left\{\begin{aligned}
        \eta\left[t_{\lb{n}}(Z_a^1),t_{\lb{n}}(z_a)\right] & ~ \text{ if } ~ z_a \in Z_a^1, \\ 
        -1 & ~ \text{ if } ~ z_a \in Z_a^2, \\
        \eta\left[t_{\ub{n}}(Z_a^3),t_{\ub{n}}(z_a)\right] & ~ \text{ if } ~ z_a \in Z_a^3.
        \end{aligned} \right.
\end{equation}     

We now describe a method to obtain the polyhedral relaxation $Q^\text{CV}_j$ using linearizations of $\betacv(z_a)$ in each of the regions $Z_a^{1}$, $Z_a^{2}$, and $Z_a^{3}$. Begin with $Q^\text{CV,1}_j=Q^\text{CV,2}_j=Q^\text{CV,3}_j=\mathbb{R}^{n_z}$. If $Z_a^1\neq\emptyset$, set $Q^\text{CV,1}_j=\mathcal{Q}^\text{CV}_\eta(t_{n^L}(Z_a^1))$. By definition, all of the constraints in $\mathcal{Q}^\text{CV}_\eta(t_{n^L}(Z_a^1))$ are linearizations of $\eta\left[t_{\lb{n}}(Z_a^1),t_{\lb{n}}(z_a)\right]$ at points $t_{\lb{n}}(z_a)$ with $z_a\in Z_a^1$. By definition of $\beta_j^\text{CV}$, these are linearizations of $\beta_j^\text{CV}$ at points in $Z_a^1$. Since $\beta_j^\text{CV}$ is convex, these must be valid underestimators. If $Z_a^2\neq\emptyset$, set $Q^\text{CV,2}_j=\{\mathbf{z}\in\mathbb{R}^{n_z}:z_j\geq-1\}$. Finally, if $Z_a^3\neq\emptyset$, set $Q^\text{CV,3}_j=\mathcal{Q}^\text{CV}_\eta(t_{n^U}(Z_a^3))$. Then, the final enclosure is given by $Q^\text{CV}\triangleq Q^\text{CV,1}_j\cap Q^\text{CV,2}_j\cap Q^\text{CV,3}_j$. Note that the number of inequalities in $Q^\text{CV}_j$ varies according to the emptiness of the respective regions.


According to \cite{ScottPhdThesis}, a valid concave overestimator of $\sin(z_a)$ for $z_a \in Z_a$ is given by $\betacc(z_a) \triangleq -\betacv(z_b)$ for $z_b \in Z_b$, where $z_b \triangleq -z_a$ and $Z_b \triangleq -Z_a$. Then, the derivation of the concave relaxation $Q^\text{CC}_j$ is analogous to $Q^\text{CV}_j$ using a simple variable transformation. The polyhedron relaxation $Q_j \supseteq \{\mbf{z} \in Z : z_j = \sin(z_a)\}$ is given by $Q^\text{CV}_j \cap Q^\text{CC}_j$.

\section{Proofs of Section \ref{sec:equivalentset}} \label{app:proofs}

\proof[Proof of Lemma \ref{lem:Aee properties}] Choose any $k\in\{1,\ldots,n_\text{e}\}$. We will show that the $k^\text{th}$ row of $\mathbf{A}_\text{ee}$ has a one on its diagonal and zeroes to the right of the diagonal. Since $k$ is arbitrary, this completes the proof.

Let $\bm{\sigma}\in\mathbb{N}^{n_\text{e}}$ be a vector containing the indices in $\mathcal{I}_\text{e}$ in ascending order, and let $j=\bm{\sigma}(k)$. Since $j\in\mathcal{I}_\text{e}$, $\alpha_j$ is given by one of the operations in \eqref{eq:operationstobesolved}. Assume $\alpha_j(z_a,z_b)=z_a+z_b$. The proofs in the other cases are analogous. By Assumption \ref{ass:intervalZ}, $Q_j$ adds the single constraint $z_j=z_a+z_b$ to $P_{\ell}$, represented as a row of $\mbf{A}_p$. Denoting this row by $\bm{\mu}\in\realset^{1\times n_z}$, we know that $\mu_j=1$, $\mu_a=\mu_b=-1$, and the remaining elements are zero. By the definition of $\mbf{V}$, $\bm{\mu}$ is the $k^\text{th}$ row of $\mbf{A}_\text{e}$. Define $[\bm{\mu}_{\text{ee}} \ \bm{\mu}_\text{er}] = \bm{\mu}\mbf{U}^{-1}$. We must show that $\mu_{\text{ee},k}=1$ and $\mu_{\text{ee},q}=0$ for all $q>k$.

By the definition of $\mbf{U}$, $z_j$ will become the $k^\text{th}$ element of $\mbf{z}_\text{e}$. Accordingly, the column permutation $\mbf{U}^{-1}$ will move $\mu_j$ to the $k^{\text{th}}$ position in $\bm{\mu}_{\text{ee}}$. Thus, $\mu_{\text{ee},k}=1$.

The only other nonzeroes in $\bm{\mu}$ are $\mu_a$ and $\mu_b$, and $a,b<j$ by Definition \ref{def:factorable}. If $a\in\mathcal{I}_\text{e}$, then $\mbf{U}$ permutes $z_a$ into $\mbf{z}_\text{e}$. However, $a<j$ implies that $z_a$ will appear before $z_j$ in the vector $\mbf{z}_\text{e}$. Accordingly, $\mbf{U}^{-1}$ will move the $\mu_a$ to some column of $\bm{\mu}_{\text{ee}}$ before $k$. Alternatively, if $a\notin\mathcal{I}_\text{e}$, then $\mbf{U}^{-1}$ permutes $\mu_a$ into some column of $\bm{\mu}_{\text{er}}$, not $\bm{\mu}_{\text{ee}}$. The same arguments hold for $b$. Thus, $\mu_{\text{ee},q}=0$ for all $q>k$.
\qed

\proof[Proof of Lemma \ref{thm:equivalentpolytope}] Choose any $\mbf{z}\in\realset^{n_z}$ and let $(\mbf{z}_\text{e},\mbf{z}_\text{r})=\mbf{Uz}$. To prove the forward implication, assume $\mbf{z} \in P_\ell \cap P_y$. By definition of $(\mbf{z}_\text{e},\mbf{z}_\text{r})$, \eqref{eq:equalitiespartition} and \eqref{eq:inequalitiespartition} hold. The first block row of \eqref{eq:equalitiespartition} implies \eqref{eq:factorelimination}, and substituting \eqref{eq:factorelimination} into \eqref{eq:equalitiespartition} and \eqref{eq:inequalitiespartition}
further gives
\begin{alignat}{1} \label{eq:czprnewequalities}
    (\mbf{A}_\text{rr} - \mbf{A}_\text{re} \mbf{A}_\text{ee}^{-1}  \mbf{A}_\text{er}) \mbf{z}_\text{r} &= \mbf{b}_\text{r} - \mbf{A}_\text{re} \mbf{A}_\text{ee}^{-1}  \mbf{b}_\text{e}, \\
 \label{eq:czprnewinequalities}
    (\mbf{H}_\text{r} - \mbf{H}_\text{e} \mbf{A}_\text{ee}^{-1}  \mbf{A}_\text{er}) \mbf{z}_\text{r} &\leq \mbf{k}_p - \mbf{H}_\text{e} \mbf{A}_\text{ee}^{-1}  \mbf{b}_\text{e}.
\end{alignat}
Therefore, $\mbf{z}_\text{r} \in \mathring{P}$ and \eqref{eq:factorelimination} holds.

To prove the reverse implication, assume $\mbf{z}_\text{r} \in \mathring{P}$ and \eqref{eq:factorelimination} holds. Since $\mbf{z}_\text{r} \in \mathring{P}$, \eqref{eq:czprnewequalities} and \eqref{eq:czprnewinequalities} hold. Moreover, rearranging \eqref{eq:factorelimination} gives $\mbf{A}_\text{er}\mbf{z}_\text{r} = \mbf{b}_\text{e} - \mbf{A}_\text{ee}\mbf{z}_\text{e}$. Substituting this result into \eqref{eq:czprnewequalities} and \eqref{eq:czprnewinequalities} gives $\mbf{A}_\text{re}\mbf{z}_\text{e} + \mbf{A}_\text{rr}\mbf{z}_\text{r} = \mbf{b}_\text{r}$ and $\mbf{H}_\text{e} \mbf{z}_\text{e} + \mbf{H}_\text{r} \mbf{z}_\text{r} \leq \mbf{k}_p$, respectively. Since $(\mbf{z}_\text{e},\mbf{z}_\text{r})=\mbf{Uz}$, these imply $\mbf{A}_p \mbf{z} = \mbf{b}_p$ and $\mbf{H}_p \mbf{z} \leq \mbf{k}_p$. Therefore, $\mbf{z} \in P_\ell \cap P_y$. \qed

\proof[Proof of Lemma \ref{lem: Zr maps into Ze}] The argument requires a new function, defined as follows. Let $[\bm{\alpha}_\text{e} \  \bm{\alpha}_\text{r}]=\mbf{U}\bm{\alpha}$ so that $\bm{\alpha}_\text{e}$ contains all elementary operations in the factorable representation of $\bm{\ell}$ with $i\in\mathcal{I}_e$ (i.e., the ones defined by \eqref{eq:operationstobesolved}). Applying the operations in $\bm{\alpha}_\text{e}$ in order of increasing $j$, analogously to Definition \ref{def:factorable}, defines a factorable function of the form $\mbf{z}_\text{e}=\mbf{h}(\mbf{z}_\text{r})$. For this function to be well-defined, we only need to ensure that each operation $z_j=\alpha_j(z_a,z_b)$ has well-defined inputs $z_a$ and $z_b$ when it is executed. If $a\notin\mathcal{I}_\text{e}$, then $z_a$ is some element of $\mbf{z}_\text{r}$ and is specified as input to $\mbf{h}$. If $a\in\mathcal{I}_\text{e}$, then $z_a$ must be the output of some other operation in $\mbf{h}$, and will be computed prior to $z_j$ since $a<j$. Since the same holds for $b$, $\mbf{h}$ is well defined.

We first show that 
\begin{equation}
\label{eq: h = linear alg solution}
    \mbf{h}(\mbf{z}_{\text{r}}) = \mbf{A}_\text{ee}^{-1} (\mbf{b}_\text{e} - \mbf{A}_\text{er} \mbf{z}_\text{r}), \quad\forall \mbf{z}_\text{r}\in Z_\text{r}.
\end{equation}
Recall that the first block row of \eqref{eq:equalitiespartition} contains the equality constraints $z_j=\alpha_j(z_a,z_b)$ for all $\alpha_j$ satisfying \eqref{eq:operationstobesolved} (i.e., $j\in \mathcal{I}_\text{e}$), and nothing more. These are exactly the operations in $\bm{\alpha}_\text{r}$ that define $\mbf{h}$. It follows that, for any fixed $\mbf{z}_\text{r}$, the vector $\mbf{z}_\text{e}=\mbf{h}(\mbf{z}_\text{r})$ is a solution to the first block row of \eqref{eq:equalitiespartition}. But since $\mbf{A}_\text{ee}$ is invertible, this solution is unique and also satisfies $\mbf{z}_\text{e}=\mbf{A}_\text{ee}^{-1} (\mbf{b}_\text{e} - \mbf{A}_\text{er} \mbf{z}_\text{r})$. Thus, \eqref{eq: h = linear alg solution} holds.

Given \eqref{eq: h = linear alg solution}, it now suffices to show that
\begin{equation}
\label{eq: h maps into in Ze}
    \mbf{h}(\mbf{z}_{\text{r}}) \in Z_\text{e}, \quad\forall \mbf{z}_\text{r}\in Z_\text{r}.
\end{equation}
Let $Z_\text{e}'$ be the interval enclosure of the image $\mbf{h}(Z_{\text{r}})$ obtained by applying interval arithmetic to the factorable representation of $\mbf{h}$, exactly as for $\bm{\ell}$ in Assumption \ref{ass:intervalZ}. Let $Z'$ denote the interval $Z_\text{r}\times Z_\text{e}'$ reordered under $\mbf{U}^{-1}$ to match the original ordering of $Z$. We will show that $Z=Z'$ by induction. Choose any $j\leq n_z$ and assume that $Z_k=Z'_k$ for all $k<j$. This is true for $j = n_x+n_w+n_v+1$ because the first $n_x+n_w+n_v$ components of both $Z$ and $Z'$ correspond to components of $Z_\text{r}$. If $j\notin\mathcal{I}_\text{e}$, then $Z_j$ and $Z_j'$ also correspond to some component of $Z_\text{r}$ and must be the same. If $j\in\mathcal{I}_\text{e}$, then, by Assumption \ref{ass:intervalZ}, 
\begin{alignat}{1}
Z_j = \alpha_j(Z_a,Z_b),
\end{alignat}
where $\alpha_j$ is evaluated in interval arithmetic. However, by the construction of $\mbf{h}$ and $Z'$, we must also have
\begin{alignat}{1}
Z_j' = \alpha_j(Z_a',Z_b').
\end{alignat}
Since $Z_a'=Z_a$ and $Z_b'=Z_b$ by the inductive hypothesis, $Z_j'=Z_j$ as well. Induction over $j$ now gives $Z'=Z$ and, upon reordering, $Z_\text{e}'=Z_\text{e}$. Since $\mbf{h}(Z_{\text{r}})\subseteq Z_\text{e}'$, \eqref{eq: h maps into in Ze} must hold. \qed

\proof[Proof of Theorem \ref{lem:newpredictionupdate}] Proposition \ref{thm:predictionupdate} shows that $\underline{X}_k\subseteq \hat{X}_k$. Thus, to prove Conclusion 1, it suffices to show that $\hat{X}_k\subseteq \hat{X}_k'$. Choose any $\mbf{x}_k\in \hat{X}_k$. By \eqref{eq:czprpredictionupdate}, there exists $\mbf{z}$ in $(\hat{X}_{k-1} \times W \times V \times \tilde{Z}) \cap P_\ell \cap P_y$ such that
\begin{equation}\label{eq: x=Ez}
\mbf{x}_k=\mbf{E}_f\mbf{z}.
\end{equation}
Let $(\mbf{z}_\text{e},\mbf{z}_\text{r})=\mbf{Uz}$. Since $\mbf{z}\in P_\ell \cap P_y$, Lemma \ref{thm:equivalentpolytope} ensures that $\mbf{z}_\text{r}\in\mathring{P}$ and \eqref{eq:factorelimination} holds. Moreover, $\mbf{z}\in \hat{X}_{k-1} \times W \times V \times \tilde{Z}$ implies that $\mbf{z}_\text{r}\in \hat{X}_{k-1} \times W \times V \times \tilde{Z}_\text{r}$. To see this, note that no $j\leq n_x+n_w+n_v$ satisfies \eqref{eq:operationstobesolved}, and so no such $j$ is in $\mathcal{I}_\text{e}$. By definition of $\mbf{U}$, it follows that the first $n_x+n_w+n_v$ elements of $\mbf{z}_\text{r}$ are the same as the first $n_x+n_w+n_v$ elements in $\mbf{z}$, and hence lie in $\hat{X}_{k-1} \times W \times V$.

Since $\mbf{z}_\text{r}\in (\hat{X}_{k-1} \times W \times V \times \tilde{Z}_\text{r})\cap\mathring{P}$, \eqref{eq:czprnewpredictionupdate} implies that
\begin{alignat}{1}
\hat{X}_k' &\ni \mbf{G}_f\mbf{z}_\text{r}+\mbf{c}_f, \\
&= (\mbf{E}_\text{r} - \mbf{E}_\text{e} \mbf{A}_\text{ee}^{-1}  \mbf{A}_\text{er})\mbf{z}_\text{r}+\mbf{E}_\text{e} \mbf{A}_\text{ee}^{-1}  \mbf{b}_\text{e}, \\
&= \mbf{E}_\text{r}\mbf{z}_\text{r} + \mbf{E}_\text{e} \mbf{A}_\text{ee}^{-1}(\mbf{b}_\text{e}-\mbf{A}_\text{er}\mbf{z}_\text{r}), \\
&= \mbf{E}_\text{r}\mbf{z}_\text{r} + \mbf{E}_\text{e} \mbf{z}_\text{e},
\end{alignat}
where the last equality follows from \eqref{eq:factorelimination}. By \eqref{eq:Efpartition} and \eqref{eq: x=Ez}, it follows that $\mbf{x}_k\in \hat{X}_k$. Since $\mbf{x}_k$ was arbitrary, $\hat{X}_k\subseteq \hat{X}_k'$.

To prove Conclusion 2, it only remains to show that $\hat{X}_k'\subseteq \hat{X}_k$. Choose any $\mbf{x}_k\in \hat{X}_k'$. By \eqref{eq:czprnewpredictionupdate}, there exists $\mbf{z}_\text{r}\in (\hat{X}_{k-1} \times W \times V \times \tilde{Z}_\text{r})\cap \mathring{P}$ such that
\begin{alignat}{1}
\mbf{x}_k&=\mbf{G}_f\mbf{z}_\text{r}+\mbf{c}_f, \\
&= (\mbf{E}_\text{r} - \mbf{E}_\text{e} \mbf{A}_\text{ee}^{-1}  \mbf{A}_\text{er})\mbf{z}_\text{r}+\mbf{E}_\text{e} \mbf{A}_\text{ee}^{-1}  \mbf{b}_\text{e}, \\
&= \mbf{E}_\text{r}\mbf{z}_\text{r} + \mbf{E}_\text{e} \mbf{A}_\text{ee}^{-1}(\mbf{b}_\text{e}-\mbf{A}_\text{er}\mbf{z}_\text{r}).
\end{alignat}
Define $\mbf{z}_\text{e}$ by \eqref{eq:factorelimination} and let $\mbf{z} = \mbf{U}^{-1}(\mbf{z}_\text{e},\mbf{z}_\text{r})$. Then,
\begin{alignat}{1}
\mbf{x}_k &= \mbf{E}_\text{r}\mbf{z}_\text{r} + \mbf{E}_\text{e} \mbf{z}_\text{e} = \mbf{E}_f\mbf{z}.
\end{alignat}
Finally, we show that $\mbf{z}\in(\hat{X}_{k-1} \times W \times V \times \tilde{Z}) \cap P_\ell \cap P_y$, since this combined with \eqref{eq:czprpredictionupdate} gives $\mbf{x}_k\in \hat{X}_k$, and hence $\hat{X}_k'\subset \hat{X}_k$.

Since $\mbf{z}_\text{r}\in\mathring{P}$ and \eqref{eq:factorelimination} holds, Lemma \ref{thm:equivalentpolytope} ensures that $\mbf{z}\in P_\ell \cap P_y$. Since $\mbf{z}_\text{r}\in \hat{X}_{k-1} \times W \times V \times \tilde{Z}_\text{r}\subseteq Z_\text{r}$, $\mbf{z}_\text{r}\in Z_\text{r}$. The hypothesis of Conclusion 2, combined with \eqref{eq:factorelimination}, now shows that $\mbf{z}_\text{e}\in Z_\text{e}$. Thus, $\mbf{z}\in Z$. Noting again that $\mbf{z}$ and $\mbf{z}_\text{r}$ share the first $n_x+n_w+n_v$ elements, it follows that $\mbf{z}\in\hat{X}_{k-1} \times W \times V \times \tilde{Z}$.
\qed

\bibliography{\bibfolder/masterthesis_bib,\bibfolder/appendices_bib,\bibfolder/UAVControl_bib,\bibfolder/BackgroundHist_bib,\bibfolder/Surveys_bib,\bibfolder/PassiveFTC_bib,\bibfolder/ActiveFTC_bib,\bibfolder/UAVFTC,\bibfolder/SetTheoretic_bib,\bibfolder/SetTheoreticFTCFDI_bib,\bibfolder/Davide_bib,\bibfolder/paperAutomatica_bib,\bibfolder/paperCDC_bib,\bibfolder/paperECC_bib,\bibfolder/paperIFAC_bib,\bibfolder/paperNonlinearMeas_bib,\bibfolder/Robotic_bib,\bibfolder/stelios_bibliography,\bibfolder/phdthesis_bib,\bibfolder/paperParameter_bib,\bibfolder/Diego_bib,\bibfolder/paperMixed_bib,\bibfolder/paperPolyRelax_bib,\bibfolder/MPC_bib}

\end{document}